\documentclass[twocolumn,prb,floatfix`]{revtex4-1}
\usepackage{graphicx}
\usepackage{epsfig}
\usepackage{verbatim}
\usepackage{bm}
\usepackage{physics}
\usepackage{times}
\usepackage{amsmath}
\usepackage{amsfonts}
\usepackage{amssymb}
\usepackage{hyperref}
\usepackage{textgreek}

\hypersetup{
     colorlinks   = true,
     linkcolor    = blue,
     citecolor    = blue
}
\usepackage[all]{hypcap}

\newcommand{\bv}[1]{\mathbf{#1}}

\begin{document}

\title{Wiedemann-Franz law and Fermi liquids}

\author{Ali Lavasani}
\author{Daniel Bulmash}
\author{Sankar Das Sarma}
\affiliation{Department of Physics, Condensed Matter Theory Center, University of Maryland, College Park, Maryland 20742, USA
and Joint Quantum Institute, University of Maryland, College Park, Maryland 20742, USA}

\begin{abstract}
  We consider in depth the applicability of the Wiedemann-Franz (WF) law, namely that the electronic thermal conductivity ($\kappa$) is proportional to the product of the absolute temperature ($T$) and the electrical conductivity ($\sigma$) in a metal with the constant of proportionality, the so-called Lorenz number $L_0$, being a materials-independent universal constant in all systems obeying the Fermi liquid (FL) paradigm.  It has been often stated that the validity (invalidity) of the WF law is the hallmark of an FL (non-Fermi-liquid (NFL)).  We consider, both in two (2D) and three (3D) dimensions, a system of conduction electrons at a finite temperature $T$ coupled to a bath of acoustic phonons and quenched impurities, ignoring effects of electron-electron interactions.  We find that the WF law is violated arbitrarily strongly with the effective Lorenz number vanishing at low temperatures as long as phonon scattering is stronger than impurity scattering. This happens both in 2D and in 3D for $T<T_{BG}$, where $T_{BG}$ is the Bloch-Gr\"uneisen temperature of the system.  In the absence of phonon scattering (or equivalently, when impurity scattering is much stronger than the phonon scattering), however, the WF law is restored at low temperatures even if the impurity scattering is mostly small angle forward scattering. Thus, strictly at $T=0$ the WF law is always valid in a FL in the presence of infinitesimal impurity scattering.  For strong phonon scattering, the WF law is restored for $T> T_{BG}$ (or the Debye temperature $T_D$, whichever is lower) as in usual metals.  At very high temperatures, thermal smearing of the Fermi surface causes the effective Lorenz number to go below $L_0$ manifesting a quantitative deviation from the WF law.  Our work establishes definitively that the uncritical association of an NFL behavior with the failure of the WF law is incorrect.
\end{abstract}

\maketitle

\section{Introduction}
In 1853, Franz and Wiedemann \cite{franz1853ueber} made the experimental discovery that the ratio of the thermal ($\kappa$) to the electrical conductivity ($\sigma$) in several metals is approximately the same at the same temperature.  Almost thirty years after this discovery, Lorenz established \cite{lorenz1881ueber} that this ratio of $\kappa$/$\sigma$ is in fact proportional to the absolute temperature $T$, and therefore $\kappa/(\sigma T)$ is a universal constant in all metals:
\begin{equation}
  \frac{\kappa}{\sigma T} = L(T)=L_0,
\end{equation}
where $L_0$, dependent only on the fundamental constants $k_B$ and electron charge $e$, is universally called the Lorenz number, given by:
\begin{equation}
L_0=\frac{\pi^2}{3} \qty(\frac{k_B}{e})^2 = 2.45\times 10^{-8}\,\text{W\textOmega}/\text{K}^2.
\end{equation}

We will call $L_0$ the ideal Lorenz number, and $L(T)$, the effective Lorenz number in case that this ratio deviates from the ideal $L_0$ value.  The finding that $\kappa/(\sigma T) = L_0$ is universally called the Wiedemann-Franz (WF) law.  In usual 3D metals, the room temperature value of $L_0$ is remarkably universal (with $L \sim L_0$ within $5\%$), making the WF law one of the most applicable defining characteristics of metallic (and hence, FL) properties \cite{kumar1993experimental}.  If $L(T)$ deviates from $L_0$ in a substantive manner, it is referred to as the failure of the WF law, which is then often attributed to the breakdown of the underlying quasiparticle picture and a failure of the Fermi liquid description for the relevant physics.   The current work is on a theoretical study of $L(T)$ in 2D and 3D metals (“Fermi liquids”) where the electron liquid is coupled to static random impurities (“disorder”) and acoustic phonons (“phonon bath”).  We show that $L(T)$, depending on the temperature and the details of the parameters describing the coupled electron-impurity-phonon system, could manifest arbitrary values of $L(T)$ with the effective Lorenz number strongly suppressed from the ideal value $L_0$ entirely within the Fermi liquid theory without invoking either a breakdown of the quasiparticle picture or the Fermi liquid paradigm.  We establish beyond any doubt that the mere inapplicability of the WF law to a metal does not necessarily imply an underlying NFL behavior, and the widespread use of the validity (invalidity) of the WF law as a smoking gun for an underlying FL (NFL) behavior is unwarranted.

Drude \cite{drude1900electron,drude19027} provided a simple classical kinetic theory for the WF law, leading to the following formula:
\begin{equation}
L(T)=\frac{3}{2} \qty(\frac{k_B}{e})^2 = 1.24 \times 10^{-8}\,\text{W\textOmega}/\text{K}^2.
\end{equation}

It is interesting that Drude's purely classical derivation of the WF law provides a result which is fortuitously within a factor of 2 of the ideal Lorenz number later derived by Sommerfeld \cite{sommerfeld1928elektronentheorie} using the appropriate quantum theory of solids. This arises from the fact that both derivations use elastic impurity scattering (e.g. quenched impurities or defects) as the driving kinetic mechanism for the electrical ($\sigma$) or thermal ($\kappa$) transport by the same carriers, consequently leading to the canceling out of the materials parameters (e.g. effective mass, carrier density) in the ratio $\kappa/\sigma$ of the system.  This has led to the belief that any lack of such a cancellation, with $L(T)$ manifesting strong temperature dependence, must automatically imply a hypothetical NFL situation where the “particles” carrying charge current and the “particles” carrying heat current are distinct, leading to the failure of the WF law. While an intrinsic separation of charge and energy transport occurring through different channels would most likely lead to a failure of the WF law (since the two transport mechanisms are then no longer kinetically connected), such a failure does not have to necessarily imply  the breakdown of the FL theory. By contrast, it has actually been known \cite{wilson1954theory} since the early days of the theory of metals that the same “particles” (namely, electrons or FL quasiparticles) could in fact carry charge and heat current through very different kinetic rates if the operational scattering mechanism is inelastic.  Since phonons provide such an inelastic scattering mechanism, in general, a coupled electron-phonon system should always violate the WF law.  The fact that ordinary metals obey the WF law at room temperatures in spite of their transport properties (both electrical and thermal) being dominated by phonons is simply a manifestation of the fact that the room temperature is a very high (very low) characteristic temperature for phonons (electrons) since the typical phonon Debye temperature $T_D$ (Fermi temperature $T_F$) for metals is $T_D\sim 100$ ($T_F\sim 50,000$)K.  In any system with high Debye temperature (much higher than room temperature), the WF law should be violated even at room temperatures (since, then, the phonon scattering will be inelastic even at room temperatures), and in all metals, the WF law is indeed strongly violated at low temperatures $T\ll T_D$ unless impurity scattering starts dominating over phonon scattering.  In a very clean metal $L(T)$ will be vanishingly small at $T\ll T_D$ as long as impurity scattering is negligible.

Thus, the validity or the failure of the WF law has little to do with NFL behavior, and is connected with the elastic or inelastic nature of the resistive scattering mechanism dominating transport in the system in the temperature regime where $L(T)$ is being measured.  In particular, in an FL coupled to impurities and phonons there are two important energy scales (assuming the Fermi temperature $T_F$ to be very high as it always is in the usual 3D metals):  $T_i$ and $T_p$.  For $T<T_i$, impurity scattering dominates, and the WF law is strictly valid by virtue of scattering by quenched impurities being elastic.  For $T_i<T<T_p$, inelastic phonon scattering dominates, and the WF law is strongly violated.  For $T>T_p$, phonon scattering is quasielastic as one enters the so-called equipartition regime, and the WF law is restored again.  Note that for strongly disordered systems, where $T_i>T_p$, WF law is obeyed at all temperatures.  These are the main theoretical results we present for both 2D and 3D metals in our work.  We note that $T_p$ may or may not be directly connected to the Debye temperature $T_D$ except that $T_p < T_D$.  In particular,  $T_p$ may be of the order of the characteristic temperature scale $T_{BG}$, where $T_{BG}$ is the so-called Bloch-Gruneisen temperature of the system associated with the energy of a phonon with wave vector equal to $2k_F$, i.e., $k_B T_{BG}=2 \hbar c_s k_F$, where $c_s$ is the phonon velocity.  In situations where $T_{BG}>T_D$ (as in normal metals), $T_p\sim T_D/3$. In our work, we consider the situation $T_{BG}<T_D$ in contrast to regular 3D normal metals where $T_D<T_{BG}$.  The reason for our choice of $T_{BG}<T_D$ is that our interest is in relatively low-density metals (e.g. cuprates) where this condition is likely to be met  by virtue of $k_F$ being small.

There are a few caveats one must keep in mind in this context.  First, when optical phonons, with a fixed energy $E_O$, are present in the system (we consider only acoustic phonons whose energy goes as $c_s q$ for phonon wave number $q$) then the definition of $T_p$ simply becomes $T_p\sim E_O$.  Thus in the presence of strong optical phonon scattering, the violation of the WF law may persist to rather high temperatures since $E_O$ could be large.  We do not consider optical phonons since they are typically absent in metals as a strong scattering mechanism.  Second, if the Fermi temperature is low (e.g. low-density systems) so that $T \sim T_F$ (an impossibility in usual metals since $T_F> 10,000$K), then the WF law is weakly violated even for $T>T_p$ because thermal Fermi surface smearing makes the system behave classically.  Third, all effects of electron-electron interaction are ignored in this work; electron-electron interaction effects on the WF law (without any phonon effects) have recently been discussed in Ref. \onlinecite{lucas2018electronic}.  Thus, our work includes disorder and phonons whereas Ref. \onlinecite{lucas2018electronic} includes disorder and electron-electron scattering effects.  Finally, the physics should be similar if phonons are replaced by some other bosonic scattering mechanisms leading to the resistivity, e.g. magnons, paramagnons or spin fluctuations.  This type of scattering, if present, should also produce the violation of the WF law at low enough temperature in clean enough systems as long as the temperature for measuring $L(T)$ is below the characteristic temperature scale for inelastic scattering by these bosonic excitations to be operational.

We mention that some of the results we present are qualitatively known. But no existing work covers the whole subject of the WF law in the context of Fermi liquids as we do in the current work, covering both 2D and 3D systems and impurity and phonon scattering.  (All our 2D results are completely new.)  We believe that it is important to have all of these results for the temperature-dependent WF law in one comprehensive paper since there seems to be much confusion in this topic.  In particular, a large fraction of the community seems to believe that the failure of the WF law (i.e. $L(T)<L_0$) is sufficient to conclude that the underlying material is a NFL with no well-defined quasiparticles.  This is simply untrue.  The violation of the WF law may or may not be a necessary condition for the NFL behavior\cite{mahajan2013non}, but an observation of such a violation most certainly is not sufficient to conclude that the relevant system is a NFL.  We refrain from reviewing the rather large literature connecting the violation of the WF law as an automatic signature for NFL behavior since our focus is entirely on a well-defined FL theory (with impurities and acoustic phonons) for the validity or not of the WF law.  There are many publications discussing the violation of the WF law in the context of putative NFL behavior, and we cite a few here as representative examples \cite{tanatar2007anisotropic,lee2017anomalously,dong2013anomalous,wakeham2011gross}, simply to emphasize the importance of the subject matter of our work where the violation of the WF law is studied entirely in the context of FL physics.

The rest of this article is organized as follows:  In Sec. \ref{sec:theoryAndResults} we provide the main theory and the associated numerical results for the calculated effective Lorenz number as a function of temperature for both 2D and 3D systems; Sec. \ref{sec:conclusion} provides extensive discussions and a conclusion putting our results in the appropriate context of the violation or not of the WF law with reference to the applicability or not of the Fermi liquid paradigm.  Five relevant appendices (\ref{app:BEF}-\ref{app:approximations}) provide the details of the theory complementing the presented results in the main text. we relegated the theoretical details to a series of appendices so that our main message (sections \ref{sec:theoryAndResults} and \ref{sec:conclusion} and the figures in the main text) can be read and understood without referring at all to the theoretical details.

\section{Theory and Results}\label{sec:theoryAndResults}

We use the standard Boltzmann kinetic theory with appropriate approximations to calculate the temperature-dependent effective Lorenz number in a FL metal in the presence of static disorder (arising from random quenched impurities) and acoustic phonons (treated within the Debye model) in the continuum long wavelength jellium model. We start with a brief review of the Boltzmann equation and the mathematical framework we used to study transport coefficients while highlighting the approximations which were employed to this end. We only present the final results in this section and leave most of the detailed and step by step calculations to the appendices (\ref{app:BEF}-\ref{app:approximations}) which should be consulted for the technical details.

\subsection{Boltzmann equation}
Let $f(\bv{r},\bv{k},t)$ denote the distribution function of electron wave packets at position $\bv{r}$ with wave vector $\bv{k}$ at time $t$. Evolution of $f(\bv{r},\bv{k},t)$ is governed by the Boltzmann equation\cite{ashcroft2005solid}:
\begin{equation}\label{equ:BE}
  \pdv{f}{t}+\dot{\bm r}\cdot \pdv{f}{\bm r}+\dot{\bm k}\cdot \pdv{f}{\bm k}=\mathcal{I}_{\bm k}\{f\},
\end{equation}
where $\dot{\bm r}$ and $\dot{\bm k}$ are given by semiclassical equations of motion\cite{xiao2010berry}:
\begin{align}
  \dot{\bm r}&=\bm v=\frac{1}{\hbar} \pdv{\varepsilon(\bm k)}{\bm k}-\dot{\bm k}\times \bm \Omega{(\bm k)}\nonumber \\
  \dot{\bm k}&=-\frac{e}{\hbar}\bm{E}(\bm r,t)-\frac{e}{\hbar c} \dot{\bm r} \times \bm{B}(\bm r,t),
\end{align}
where $\varepsilon(k)$ is the band dispersion, $\Omega(\bm k)$ is the Berry curvature and $\bm{E}$ ($\bm{B}$) is the external electric (magnetic) field.  For the parabolic band dispersion which we assume in this paper, $\Omega(\bm k)=0$.  We also set $\bm B$ to zero since we are  only interested in the zero field longitudinal conductivity.

$\mathcal{I}_{\bm k}\{f\}$ in Eq.\eqref{equ:BE} is the collision integral given by
\begin{align}
\label{eqn:collisionIntegral}
\mathcal{I}_{\bm k}\{f\}=-\int \frac{\dd^3 k'}{(2\pi)^3}[&S(\bm k,\bm k')f_{\bm k}(1-f_{\bm k'})\nonumber\\
&+S(\bm k',\bm k)f_{\bm k'}(1-f_{\bm k})],
\end{align}
where $S(\bm k,\bm k')$ is the differential scattering rate from state $\bm k$ to state $\bm k'$ and can be computed using Fermi's golden rule for various scattering mechanisms.

We write $f$ as
\begin{equation*}
  f(\bm r,\bm k, t)=f_0(\bm r,\bm k, t)+\delta f(\bm r, \bm k, t),
\end{equation*}
where $f_0$ is the distribution function of fermions in local equilibrium, given by the Dirac distribution
\begin{equation*}
  f_0(\bm{r},\bm{k},t)=\qty[\exp(\frac{\varepsilon(\bv k)-\mu(\bm r,t)}{k_B T(\bm r,t)})+1]^{-1},
\end{equation*}
and $\delta f$ is the deviation from that. If we plug in this form into the Boltzmann equation and only keep terms of first order in external fields and temperature gradient, we arrive at the linearized Boltzmann equation
\begin{equation}\label{linbol}
\pdv{\delta f_{\bv k}}{t}+\bm v \cdot \qty[e \bm{\mathcal{E}}+\frac{\varepsilon(\bv{k})-\mu}{T}\grad T]\qty(-\pdv{f_0}{\varepsilon})=\mathcal{I}_{\bm k}\{f_0+\delta f\},
\end{equation}
where $\bm {\mathcal{E}}=\bm E+\grad{\mu}/e$ is the electrochemical force and $\delta f_{\bm k}=\delta f(\bm r, \bm k, t)$. We have assumed that temperature and electric field are both slowly varying in space. Since our interest is in linear response, we will work with the linearized Boltzmann equation in the rest of this article.

We are interested in the steady state solution and hence the first term in Eq.\eqref{linbol} can be dropped. In the linear response regime, considering the symmetries of the problem, the following ansatz can be used to solve the linearized Boltzmann equation\cite{wilson1954theory}:
\begin{align}\label{ansatz}
  \delta f_{\bv k}= &\tau_\sigma(\varepsilon,T)\,e\, \bm v \cdot \bm{\mathcal{E}}~\qty(\pdv{f_0}{\varepsilon})\nonumber \\
  &+\tau_\kappa(\varepsilon,T)\frac{\varepsilon(\bv{k})-\mu}{T}\bm v \cdot \grad T \qty(\pdv{f_0}{\varepsilon}),
\end{align}
  where $\tau_\sigma$ and $\tau_\kappa$ are generally unknown functions which are generically distinct (hence allowing for the generic possibility of a failure of the WF law). We call $\tau_\sigma$ and $\tau_\kappa$ electrical and thermal relaxation times respectively. As we will mention shortly, whenever the scattering mechanism is elastic, these two relaxation times become equal, leading necessarily to the WF law.  By contrast, for inelastic scattering, thermal and electrical relaxation times could be completely different, thus leading to a total failure of the WF law independent of the validity or not of the FL paradigm.  The key for the validity (or not) of the WF law is the elastic or inelastic nature of carrier scattering, and not the FL or NFL nature of the underlying electron system.  Whenever transport is dominated by inelastic scattering (e.g. low temperatures $T\ll T_D$ for acoustic phonons), the corresponding scattering mechanism may strongly violate the WF law.

Some formal details of the Boltzmann transport theory are provided in Appendix \ref{app:BEF} for completeness.

\subsection{Transport coefficients}
For a given external electric field and temperature gradient in the linear response regime, electrical current $\bm{J}_e$ and  thermal current $\bm{J}_q$ would be
\begin{equation}
  \begin{pmatrix}
    \bm{J}_e\\
    \bm{J}_q
  \end{pmatrix}=
  \begin{pmatrix}
    L_{EE} & L_{ET}\\
    L_{TE} & L_{TT}
  \end{pmatrix}
  \begin{pmatrix}
    \bm{ \mathcal{E}}\\
    \grad T
  \end{pmatrix}.
  \label{equ:src_cur}
\end{equation}

Once the relaxation times appearing in the ansatz \eqref{ansatz} have been calculated, transport coefficients can be obtained using the following expressions:
\begin{align}\label{tc}
  L_{EE}&=e^2 \int_{-\infty}^{+\infty}\dd \varepsilon~ \qty(-\pdv{f_0}{\varepsilon})~ D(\varepsilon)\, v_x^2(\varepsilon) \tau_\sigma(\varepsilon)\nonumber, \\
  L_{ET}&=\frac{e}{T} \int_{-\infty}^{+\infty}\dd \varepsilon  ~\qty(-\pdv{f_0}{\varepsilon})~ (\varepsilon-\mu) ~D(\varepsilon)\, v_x^2(\varepsilon) \tau_\kappa(\varepsilon)\nonumber,\\
  L_{TE}&=-e\int_{-\infty}^{+\infty}\dd \varepsilon  ~\qty(-\pdv{f_0}{\varepsilon})~ (\varepsilon-\mu) ~D(\varepsilon)\, v_x^2(\varepsilon) \tau_\sigma(\varepsilon)\nonumber,\\
  L_{TT}&=\frac{-1}{T} \int_{-\infty}^{+\infty}\dd \varepsilon ~ \qty(-\pdv{f_0}{\varepsilon})~ (\varepsilon-\mu)^2 ~D(\varepsilon)\, v_x^2(\varepsilon) \tau_\kappa(\varepsilon) \nonumber,\\
\end{align}
where $D(\varepsilon)$ is the density of states at energy $\varepsilon$.

Electrical conductivity $\sigma$ is simply the $L_{EE}$ coefficient. The thermal conductivity $\kappa$, is defined such that $\bm{J}_q=-\kappa \grad T$ when $\bm{J}_e=0$, and with a little bit of algebra turns out to be $\kappa=-\qty(L_{TT}-\frac{L_{TE}L_{ET}}{L_{EE}})$.

In the following sections, we study transport properties of FL in the presence of two different scattering mechanisms; electron-impurity scattering and electron-phonon scattering. First, we consider each scattering mechanism separately and then we study their combined effect.

\subsection{Impurity scattering}
\begin{figure*}
  \includegraphics[width=\textwidth]{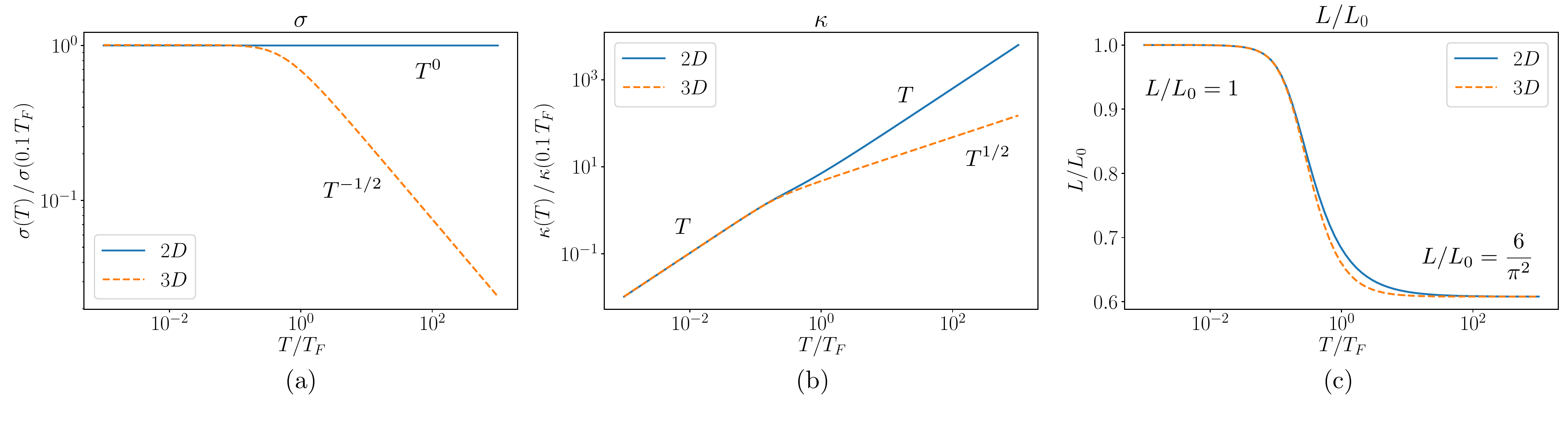}
  \caption{(a) electrical conductivity versus temperature for free electron model in presence of random impurity. (b) thermal conductivity for the same system. (c) the Lorenz ratio $L/L_0$ for the same system. In all figures, solid line and dashed line correspond to 2D and 3D systems respectively. Note that electrical and thermal conductance has been plotted relative to their value at $T=0.1\, T_F$ to make them dimensionless.  }
  \label{fig:imp}
\end{figure*}
We consider the model of randomly distributed impurities with short range potential,
\begin{equation}
  V_\text{imp}(\bv{r})=u_0\, \delta(\bv{r}),
\end{equation}
where $u_0$ is some constant characterizing the scattering strength. The impurities are assumed to be fixed and hence the scattering would be elastic. The differential scattering rate can be calculated by using Fermi's golden rule and averaging over impurity locations:
\begin{equation}\label{eqn:impurityS}
  S(\bv k ,\bv k')=\frac{2\pi}{\hbar} n_\text{imp}\, u_0^2\, \delta(\varepsilon(\bv k) - \varepsilon(\bv k')),
\end{equation}
where $n_{\text{imp}}$ corresponds to the number density of impurities.
Due to the elasticity of scattering, the ansatz in Eq.\eqref{ansatz} makes it simple to find an exact solution of the linearized Boltzmann equation for any arbitrary external electric field and temperature gradient.

By plugging the ansatz in Eq.\eqref{ansatz} into Eq.\eqref{linbol}, we can find an explicit closed form for the relaxation times. As is shown in Appendix \ref{app:impurity}, due to the elasticity of scattering, thermal and electrical relaxation times are equal and can be determined from the following integral,
\begin{equation}
  \tau(\varepsilon)^{-1}=\int \frac{\dd^d \vb{k}'}{(2\pi)^d}(1-\cos(\theta))S(\bv k ,\bv k'),
\end{equation}
where $\theta$ is the so-called scattering angle between $\bv k$ and $\bv k'$. The integral can be carried out and we get a temperature independent relaxation time with different energy dependence in 2D and 3D (we note that there could be temperature dependence if somehow the impurity potential  $u_0$ itself has temperature dependence, a possibility we ignore in the current work):
\begin{equation}\label{tau_imp}
  \tau(\varepsilon)\propto
  \begin{cases}
    \varepsilon^{0}\quad &\text{2D}\\
    \varepsilon^{-1/2}  &\text{3D}
  \end{cases}
\end{equation}
The constant of proportionality depends on the parameters of the system and can be found in Appendix \ref{app:impurity}. Having computed the relaxation time, one can  obtain the electrical and thermal conductivities using  Eq.\eqref{tc}. The calculated results are shown using dimensionless units in Figs.\ref{fig:imp}(a) and (b). As is clear from the plots, the only temperature scale that appears in this model is the Fermi temperature defining the zero point energy of the noninteracting electrons. For $T \ll T_F$, the electrical conductivity does not have any temperature dependence and the thermal conductivity is linear in $T$:
\begin{equation}
  \sigma(T) \propto T^0, \qquad \kappa(T)\propto T.
\end{equation}
The WF law is obeyed in this regime (see Fig. \ref{fig:imp}c):
\begin{equation}
  \frac{L}{L_0}=1 \qquad (T\ll T_F).
\end{equation}
On the other hand, for $T\gg T_F$, we get the following temperature scalings which differ based on the spatial dimension:
\begin{equation}
  \sigma(T)\propto
  \begin{cases}
    T^{0}\quad &\text{2D}\\
    T^{-1/2}  &\text{3D}
  \end{cases}
  ,\qquad \kappa(T)\propto
  \begin{cases}
    T\quad &\text{2D}\\
    T^{1/2}  &\text{3D}.
  \end{cases}
\end{equation}
System parameters still cancel out in the $\kappa/\sigma$ ratio in this regime, which is related to the fact that the energy and charge currents both relax with the same rate. However, smearing of the Fermi surface at $T \gg T_F$ causes the Lorenz ratio  $L/L_0$ to deviate from $1$, approaching $\frac{6}{\pi^2}<1$ as is shown in Fig.\ref{fig:imp}c. The full expressions for $\sigma(T)$ and $\kappa(T)$ as well as details of the calculation can be found in Appendix \ref{app:impurity}.

Thus, a modified WF law still applies for elastic impurity scattering at very high temperatures with a suppressed effective Lorenz number $L<L_0$.  This is of possible experimental relevance in low density metallic systems where the $T>T_F$ regime may be attainable.  Obviously, this high-T result is of no relevance to normal metals where $T_F \sim 50,000$K.

\subsection{Phonon scattering}
\label{sec:aph}

In this section we consider the effect of electron-phonon scattering on transport coefficients. Other than the electron-phonon interaction, we do not incorporate any other lattice effect into our model and work in the continuum approximation. In particular, we assume a parabolic energy dispersion for the electrons and ignore Umklapp scattering. We also neglect phonon drag. Finally, since our interest is strictly in the behavior of the Fermi liquid itself, we calculate only the electronic contribution to the thermal conductivity.

The electron-phonon interaction is given by the so-called deformation potential model:
\begin{equation}
  H_{\text{e-ph}}=\frac{1}{\sqrt{V}}\sum_{k,k',q} \qty[\sqrt{\frac{\hbar D^2}{2 \rho \omega_q}}\, q](a_q+a^\dagger_{-q})~{c_k}^\dagger c_{k'} ~\delta_{k-k'-q,0}
\end{equation}
where $a^\dagger$  and $c^\dagger$ are phonon and electron creation operators respectively. Here $D$ is the deformation potential strength, $\rho$ is the ion density, and $\omega_{\vb q}$ corresponds to the energy of a phonon with momentum $\vb q$, which, for acoustic phonons, is given as:
\begin{equation}\label{aphDispersion}
  \omega_q=c_s q
\end{equation}
with $c_s$ the speed of sound. The corresponding scattering rate, obtained from Fermi's golden rule, is:
\begin{align}
   S(\bv k, \bv k')=&\frac{2\pi}{\hbar}\qty(\frac{\hbar}{2 \rho \omega_{q}} D^2 q^2)[N_{q} ~\delta(\epsilon_k-\epsilon_{k'}+\hbar \omega_q) \nonumber \\
  & +(N_q+1)~\delta(\epsilon_k-\epsilon_{k'}-\hbar \omega_q)],
\end{align}
where
\begin{equation}\label{equ:momentumConservation}
  \bv q=\bv k - \bv k',
\end{equation}
and $N_q$ is the phonon distribution function. Since all calculations are carried out to the leading order in external fields, $N_q$ can be replaced by the equilibrium Bose-Einstein distribution function,
\begin{equation}
  N_q=\frac{1}{e^{\beta\hbar\omega_q}-1}.
\end{equation}

In contrast to the elastic case, the linearized Boltzmann equation cannot be solved exactly here because inelastic electron-phonon scattering couples the distribution function at one energy to the distribution function at another energy. To arrive at a closed form for the relaxation times, we use the relaxation time approximation, discussed in detail in Appendix \ref{app:phonon}; this uncontrolled approximation assumes that the relaxation time changes slowly enough as a function of energy that, within the collision integral, $\tau(\varepsilon) \approx \tau(\varepsilon')$. Importantly, We find that the relaxation times for electrical transport and thermal transport are generally different.

Note that we are not getting into the discussion of whether a relaxation time approximation is valid here or not as we are simply and uncritically assuming it to apply. (See Appendix \ref{app:phonon} and \ref{app:approximations} for more details.) One can of course solve the linearized Boltzmann integral equation directly numerically without assuming the relaxation time approximation (which may indeed be necessary if one is interested in a quantitative comparison with experimental results), but such a completely numerical calculation would destroy the whole purpose of our work since we are then unable to make general comments about the validity or not of the WF law.  Assuming the existence of a relaxation time (albeit possibly different ones for charge and heat transport) enables us to make considerable analytical progress without losing generality (but sacrificing quantitative accuracy).

\begin{figure*}
  \includegraphics[width=\textwidth]{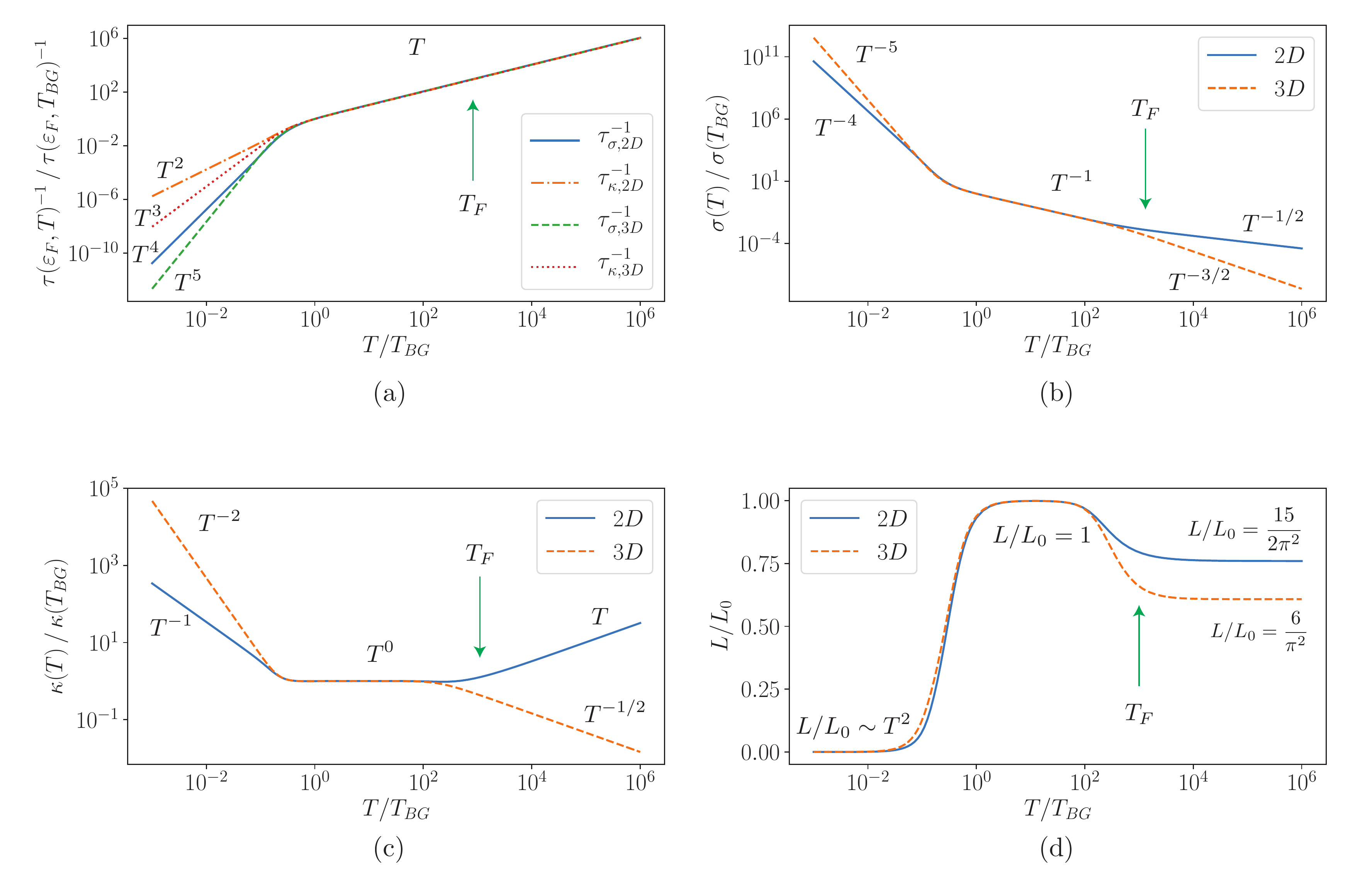}
  \caption{(a) Scattering rate $\tau^{-1}$ versus temperature for thermal and electrical transport. (b) Electrical conductivity $\sigma$ versus temperature. (c) Thermal conductivity $\kappa$ versus temperature. (d) Lorenz ratio $L/L_0$ versus temperature. In all graphs, solid and dashed lines correspond to 2D and 3D systems respectively. Relaxation rates and transport coefficients are plotted relative to their values at $T=T_{BG}$ to make them dimensionless. The Fermi energy is chosen such that $T_F=10^3 T_{BG}$}
  \label{fig:aph}
\end{figure*}

To get the relaxation time for charge transport, we use the ansatz in Eq.\eqref{ansatz} with $\grad T$ set to zero. Assuming $T \ll T_F$ and using the relaxation time approximation, we get:
\begin{align}\label{equ:tau_sigma}
  \tau_{\sigma}(\varepsilon)^{-1}=\int \frac{\dd^d \vb{k}'}{(2\pi)^d}\frac{1-f(\varepsilon')}{1-f(\varepsilon)}(1-\cos(\theta))S(\bv k ,\bv k').
\end{align}

On the other hand, to obtain the thermal relaxation time $\tau_\kappa$, we set electrochemical force $\bm {\mathcal{E}}$ to zero and solve the linearized Boltzmann equation in the presence of a non-zero temperature gradient $\grad T$.  By comparing to a more direct (but more complicated) calculation, we show in Appendix \ref{app:phonon} that the universal features of the thermal relaxation time can be approximately captured by the same expression as in Eq.\eqref{equ:tau_sigma}, but by simply dropping the $(1-\cos(\theta))$ ``forward-scattering" suppression factor in the integral\cite{ziman1979principles}:
\begin{align}\label{equ:tau_gamma}
  \tau_{\kappa}(\varepsilon)^{-1}=\int \frac{\dd^d \vb{k}'}{(2\pi)^d}\frac{1-f(\varepsilon')}{1-f(\varepsilon)}S(\bv k ,\bv k').
\end{align}
Intuitively, one can understand this difference between $\tau_\sigma$ and $\tau_\kappa$ by noting that different types of scatterings are responsible for relaxing charge and heat currents. Note that forward scattering events can not change the charge current significantly; a fact that explains the $(1-\cos(\theta))$ factor in Eq.\eqref{equ:tau_sigma}. On the other hand, the thermal current which is caused by the imbalance in the populations of electrons and holes in the vicinity of the Fermi surface (see Eq.\eqref{ansatz}), can be relaxed effectively by forward scattering which justifies the absence of $(1-\cos(\theta))$ in Eq.~\eqref{equ:tau_gamma}\cite{ziman1979principles,mahan2013many}. Thus, in the presence of inelastic scattering processes, it is sensible to include (exclude) the forward scattering suppression factor for charge (heat) current within the relaxation time approximation. We remark that when the scattering is elastic, e.g. the impurity scattering,  backscattering is the only relaxation mechanism for the thermal current (as well as the charge current) and hence the $(1-\cos(\theta))$ factor cannot be dropped in that case. Detailed calculation of relaxation times is left to the Appendix \ref{app:phonon} whereas in Appendix \ref{app:approximations} we provide a detailed discussion of the relaxation time approximation (RTA) in this context. In particular, we discuss how using different relaxation times could result in violations of the Onsager relations.

With electron-phonon scattering present, both thermal and electrical relaxation times become functions of temperature. Fig.~\ref{fig:aph}(a) shows their temperature dependence over a range of temperatures which covers multiple orders of magnitude. There are two different regimes, with the crossover occurring roughly at $k_B T_{BG}=2\hbar c_s k_F$. Note that because $T$ is much less than $T_F$, only electrons in the vicinity of the Fermi surface participate in charge and energy transport. This in turn means that momentum transfer in a scattering event is bounded by $\sim 2 k_F$. Therefore, $k_B T_{BG}$ represents an upper bound on the energy of phonons contributing to the current relaxation. We assume that $T_{BG} < T_D$, where $T_D$ is the Debye temperature; since $T_D$ only enters the problem as another upper bound on phonon energy, $T_D$ is not an important scale in the problem. We note that textbooks usually do not emphasize the importance of $T_{BG}$ in the context of metallic transport limited by phonon scattering\cite{ashcroft2005solid} since for normal metals, $k_F$ is typically very large (since normal metals have very high carrier density) leading to $T_{BG}>T_D$, and hence the phonon energy cut off for normal metals is invariably $T_D$ and not $T_{BG}$.  Since the results for $T_D$ being the cut off are already available in the literature, we focus on the relatively low density situation where $T_{BG}<T_D$, leading to $T_{BG}$ being the appropriate phonon cut off.  For high density regular metals, $T_{BG}$ in our results should simply be replaced by $T_D$; basically, the phonon cut off is either $T_{BG}$ or $T_D$ depending on whichever is smaller for the particular material.

For $T\gg T_{BG}$, thermal and electrical relaxation rates become equal and scale linearly with temperature, independent of the spacial dimension (this is the so-called phonon equipartition regime where the acoustic phonon scattering is essentially quasi-elastic in metals):
\begin{equation}
  \tau_\sigma(T)^{-1}=\tau_\kappa^{-1}(T)\propto T.
\end{equation}
But for $T\ll T_{BG}$, charge current relaxes at a smaller rate than the thermal current(see Fig.~\ref{fig:aph}a):
\begin{equation}\label{equ:tau_ll}
  \tau^{-1}_\sigma(T)\propto
  \begin{cases}
    T^{4}\quad &\text{2D}\\
    T^{5}  &\text{3D}
  \end{cases}
  ,\qquad \tau^{-1}_\kappa(T)\propto
  \begin{cases}
    T^2\quad &\text{2D}\\
    T^{3}  &\text{3D}.
  \end{cases}
\end{equation}
This can be understood intuitively as follows. Note that charge and thermal current carried by an electron can be roughly written as its charge and energy respectively times its velocity:
\begin{equation}
  \bm j_e \sim e\, \bv v, \qquad  \bm j_q \sim \varepsilon \, \bv v
\end{equation}
where $\varepsilon$ denotes the energy relative to the chemical potential. Now, the only way a scattering event can relax the charge current is by changing the electron's velocity vector since its charge is strictly conserved. This is the reason that backscattering is the most effective way to relax the charge current.  On the other hand, the thermal current can be relaxed by either changing the electron's velocity or just changing its energy when inelastic scattering processes are operational. When $T_{BG} \ll T$, the scattering becomes quasi-elastic since phonon's energy is bounded by $T_{BG}$ so a single scattering event can only change an electron's energy by a small fraction of its average value $\varepsilon \sim T$. As a result both thermal and charge current relaxations are dominated by backscattering events for $T\gg T_{BG}$ and hence the two relaxation times become equal.  On the other hand for $T \ll T_{BG}$, backscattering is exponentially suppressed due to the Bose distribution function whereas thermal current can now be relaxed effectively by changing the electrons' energy. As a result, charge current relaxes much more slowly than the thermal current. As can be seen from Eq.\eqref{equ:tau_ll}, for both 2D and 3D we have:
\begin{equation}\label{equ:tau_ratio}
  \frac{\tau^{-1}_\sigma}{\tau^{-1}_\kappa} \propto T^2
\end{equation}
and this ratio goes to zero as as $T$ goes to $0$. As one would expect, due to different relaxation times, the WF law will no longer be obeyed in this regime. In fact, if the electron-phonon scattering is the only resistive mechanism (i.e. very clean metals with no impurities), then this WF law violation is arbitrarily strong since $L(T)$ vanishes as $T$ approaches zero, making $L(T)\ll L_0$ even for a simple FL!  Thus, all one needs is a very clean FL to see an arbitrarily strong violation of the WF law at low temperatures.

Note that to violate the WF law, suppressing backward scattering just by itself is insufficient; it is crucial to allow for inelasticity. To see this clearly, one may consider the simple case of random impurities, but with a scattering potential which suppresses scattering events with momentum transfers larger than some constant value $q_0$. This situation actually arises %in bilayer systems
in, for example, delta-doped two-dimensional electron gases
where the mobile carriers and impurities live in different layers; $q_0$ is then given by $2\pi \hbar/z_0$ where $z_0$ is the layer separation\cite{DasSarma1985}. Although backscattering is suppressed in these systems for $q_0 < k_F$, WF law is still obeyed due to the elastic nature of the scatterings. This problem is studied in detail in Appendix \ref{app:forwardImpurity}, clearly establishing that the absence of backscattering by itself, without any inelasticity, does not lead to any violation of the WF law.

By plugging in the thermal and charge relaxation times into Eq.~\eqref{tc}, electrical and thermal conductivity, $\sigma(T)$ and $\kappa(T)$ can be obtained (See Appendix \ref{app:phonon}). The result is plotted in  Figs.~\ref{fig:aph}b and c. Although we expect that the relaxation time approximation to become less valid as one approaches $T\sim T_F$, we have used the same expression for the relaxation time throughout all temperature scales, even for $T\gg T_F$. Hence, our results at $T \gg T_F$ should not be interpreted quantitatively but are rather only intended to demonstrate qualitatively that at high temperatures, the Lorenz number would deviate from unity due to the smearing of the Fermi surface.

In the equipartition regime where $T_{BG}\ll T \ll T_F$, thermal conductivity is independent of temperature whereas electrical conductivity decreases as $1/T$ (the well known linear growth of resistivity in metals due to phonons\cite{mott1958theory}), both independent of dimension:
\begin{equation}
  \sigma(T) \propto T^{-1}, \qquad \kappa(T)\propto T^0.
\end{equation}
The WF law is obeyed in this regime due to the quasi-elastic nature of scatterings,
\begin{equation}
  \frac{L}{L_0} = 1\qquad (T_{BG}\ll T \ll T_F).
\end{equation}
On the other hand for $T \ll T_{BG}$, temperature scalings of $\sigma$ and $\kappa$ become dimension dependent:
\begin{equation}
  \sigma(T)\propto
  \begin{cases}
    T^{-4}\quad &\text{2D}\\
    T^{-5}  &\text{3D}
  \end{cases}
  ,\qquad \kappa(T)\propto
  \begin{cases}
    T^{-1}\quad &\text{2D}\\
    T^{-2}  &\text{3D},
  \end{cases}
\end{equation}
recovering the well known $T^5$ scaling for electrical resistivity (known as the Bloch-Gr\"uneisen formula \cite{Bloch1930,Gruneisen}) and $T^2$ scaling for thermal resistivity ($1/\kappa$) in 3D metals\cite{Kroll1933}. As one also expects from Eq.\eqref{equ:tau_ratio}, WF law is parametrically violated in this regime with the Lorenz ratio vanishing as
\begin{equation}
  \frac{L}{L_0}\propto T^2 \qquad (T\ll T_{BG}),
\end{equation}
for small temperatures. The calculated Lorenz ratio throughout all three temperature regimes is plotted in Fig.\ref{fig:aph}d for both 2D and 3D systems.

We remark that even though the violation of the WF law in this system can be traced back to different relaxation times (see Eq.\eqref{equ:tau_ratio}), the WF law in FL could still be violated at low temperatures when both energy and charge transport are characterized by a single relaxation time. In Appendix \ref{app:wf_single}, we provide results under the assumption of a single relaxation time controlling both charge and heat currents (which is not valid generally for phonon scattering). This violation is, however, not arbitrarily strong as $L(T)/L_0$ eventually becomes a constant ($<1$) for $T\ll T_{BG}$.

\subsection{Impurity \& phonon scattering }
Finally we consider the case where both scattering mechanisms (impurity scattering and electron phonon scattering) are present.
With our current approximations, the scattering rates add, which leads to
\begin{equation}\label{mr}
  \frac{1}{\tau_{\text{total}}(\varepsilon)}=\frac{1}{\tau_\text{imp}(\varepsilon)}+\frac{1}{\tau_\text{el-ph}(\varepsilon)}
\end{equation}
Where $\tau_\text{imp}^{-1}$ and $\tau_\text{el-ph}^{-1}$ correspond to the scattering rate from impurities and phonons respectively. Using Eq.\eqref{tc} it is straightforward to compute transport coefficients and hence the Lorenz ratio.

\begin{figure}
  \includegraphics[width=\columnwidth]{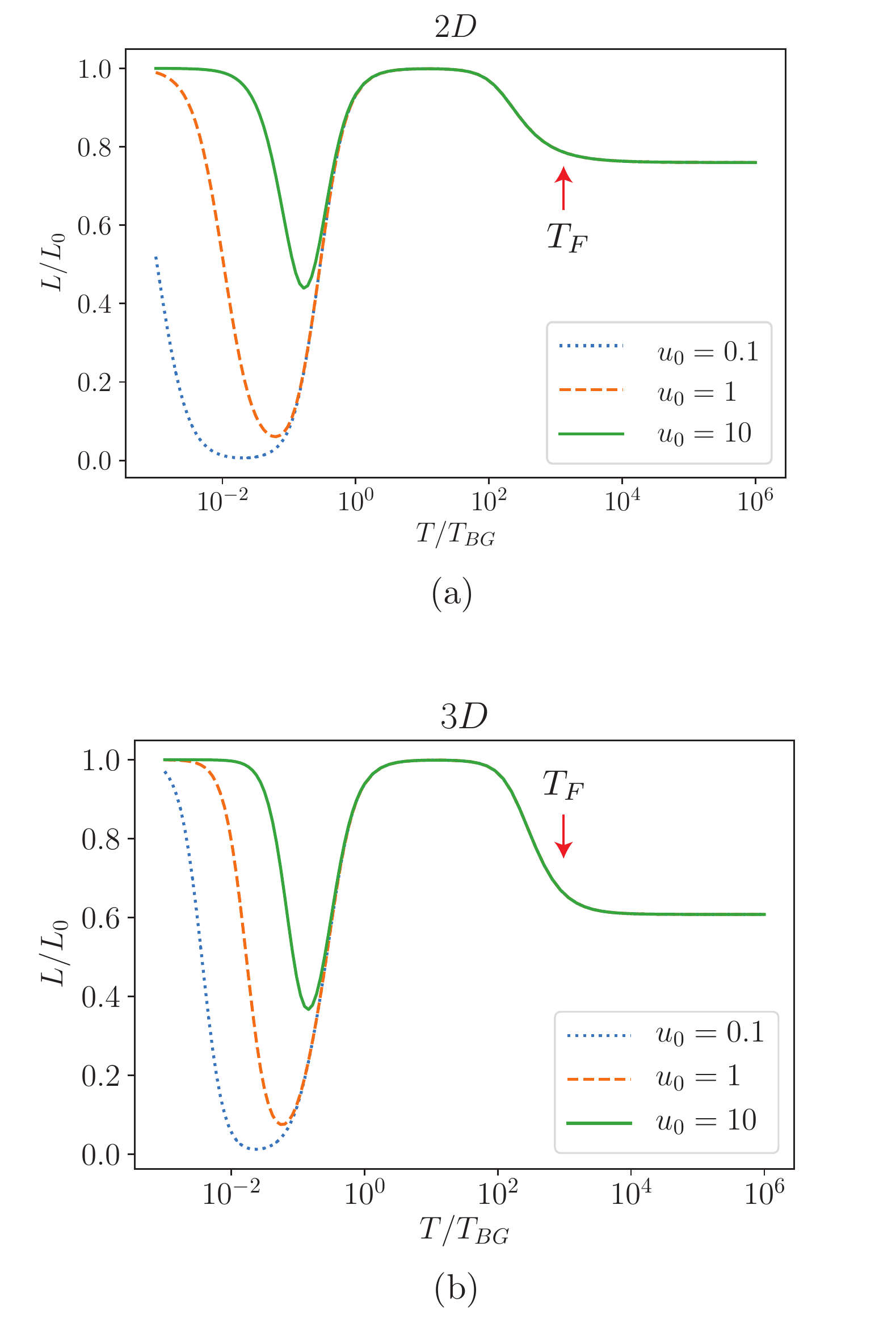}
  \caption{(a) Lorenz ratio versus temperature for 2D electron gas in presence of impurity and phonons for three different impurity densities. (b) The same results for 3D system. For both systems $T_F=10^3\,T_{BG}$. }
  \label{fig:both}
\end{figure}

The Lorenz ratio for 2- and 3- dimensional systems for three different impurity coupling strength is plotted in Fig.\ref{fig:both}. As can be seen in the figure, at low enough temperatures, the WF law is always obeyed due to the fact that eventually impurity scattering dominates transport because phonons will no longer be thermally excited at sufficiently low temperatures (but the impurity scattering is present even at $T=0$). However, as one increases the temperature, phonon scattering become stronger and, at some intermediate temperature scale $T_i$, eventually overcomes impurity scattering as the dominant scattering mechanism. At $T_i$, the Lorenz ratio starts to deviate from unity. Clearly, $T_i$ is not universal and depends on the specific parameters of the sample. WF law is violated  for $T_i \ll T \ll T_{BG}$, but is recovered again for $T_{BG}\ll T \ll T_F$ where the system is in the equipartition regime. Thus, the violation or not of the WF law in a pure Fermi liquid depends entirely on the details of the electron-phonon and electron-impurity scattering.  As long as the impurity scattering is weak (i.e. in a relatively clean metal), the WF law will be violated strongly for $T_i \ll T\ll T_{BG}$, where $T_i$ is determined by the strength of the impurity scattering in the system.  For a hypothetical absolutely clean Fermi liquid $T_i=0$ and the WF law is violated infinitely strongly at low temperatures ($T\ll T_{BG}$) as $L(T)$ vanishes.  At ``high" temperatures ($T\gg T_{BG}$), however, the WF law is strictly obeyed since phonon scattering becomes quasi-elastic in this equipartition regime (where the electrical resistivity of the metal should manifest the well-known linear-in-$T$ metallic behavior due to electron-phonon scattering) even if impurity scattering is weak (i.e. even for a very clean metal)  Thus NFL is by no means necessary for violating the WF law; the violation is automatic in a standard FL at low temperatures provided the system is clean.  Note that in relatively dirty impure metals, we may have $T_i>T_{BG}$, leading to the WF law being obeyed at all temperatures.

\section{Discussion and Conclusion}\label{sec:conclusion}
We have revisited the old topic of the Wiedemann-Franz law in 2D and 3D electron liquids interacting with quenched impurities and acoustic phonons, providing detailed results for the temperature dependent effective Lorenz number (defined as the ratio of $\kappa/\sigma\,T$) from $T=0$ to $T=T_F$.  We neglect effects of electron-electron interaction, and use the Boltzmann transport theory for obtaining the results.  Our main qualitative finding is that the WF law is strongly violated at ‘low’ temperatures ($T<T_{BG}$) in clean Fermi liquids coupled to phonons.  While most of our presented theoretical results are new, the main conclusion is neither surprising nor unknown, but seems to have been forgotten or overlooked in the currently active research on non-Fermi-liquid physics where one often associates the failure of the WF law as synonymous with the breakdown of the Fermi liquid paradigm.  Of course, in a narrow technical sense a coupled electron-phonon system is not a precise Fermi liquid \cite{engelsberg1963coupled,prange1964transport} because the interacting system has additional structures associated with phonon coupling with no analogs in the corresponding noninteracting Fermi gas, so perhaps the statement that the failure of the WF law may imply an NFL behavior is strictly speaking applicable to our system.  But the WF law is in fact restored in the coupled electron-phonon system, as our results clearly show (and as is well-known), at higher temperatures ($T>T_{BG}$), and indeed normal metals all obey the WF law rather accurately at room temperatures in spite of being a coupled electron-phonon system. In any case, electron-phonon coupling is generic in all electronic materials, and branding such a common system to be an NFL simply because it strongly violates the WF law at low temperatures is not a meaningful advance.

We show that in the presence of both impurity and phonon scattering, both 2D and 3D metals have four distinct temperature regimes, in principle, with respect to the WF law behavior:  At very low temperatures, where impurity scattering dominates over phonon scattering (with the electrical resistivity not manifesting any temperature dependence) the WF law is obeyed; at low to intermediate temperatures (but $T<T_{BG}$), where phonon scattering is stronger than impurity scattering (e.g. in clean systems) and the phonon-induced electrical resistivity shows the strong Bloch-Gr\"uneisen temperature dependence, the WF law is strongly violated due to the inelastic nature of phonon scattering; at intermediate to high temperatures, where phonons are in the equipartition regime with phonon scattering being quasielastic in nature with the electrical resistivity reflecting a linear-in-T resistivity (as normal metals always do at room temperatures), the WF law is obeyed; and finally, at very high temperatures, where $T$ approaches $T_F$, the system becomes nondegenerate and the WF law is violated weakly with the effective Lorenz number being somewhat smaller than the ideal Lorenz number. The existence of these four distinct regimes is generic both in 2D and 3D, but it is quite possible that a real material may not manifest all of these distinct regimes, depending on the parameter values controlling the various scattering strengths.  For example, a normal 3D metal with $T_F \sim 50,000K$  obviously never manifests the nondegeneracy effect of $L(T)<L_0$ at high temperatures, but 2D and 3D doped semiconductors, with $T_F \sim 100K$ or less, should have a room temperature Lorenz number typically smaller than the ideal Lorenz number by virtue of the Fermi surface nondegeneracy effect.  If the impurity scattering is strong (i.e. relatively dirty systems), then it is possible that the WF law is obeyed at all temperatures with the impurity scattering dominating at low to intermediate temperatures (up to $T_{BG}$ or above) and then quasi-elastic phonon scattering taking over at intermediate to high temperatures ($T>T_{BG}$).  This appears to be the situation in most normal metals where any violation of the WF law is uncommon at any temperature and requires very clean samples.  In fact, this accidental universal applicability of the WF law in normal 3D metals, by virtue of the overlapping elastic phonon and impurity scattering at intermediate temperatures, is what may have led to the misleading characterization of the validity or not of the WF law as implying the validity or not of the FL theory.  In fact, an arbitrarily clean FL metal would violate the WF law at arbitrarily low temperature with $L(T)/L_0 \sim T^2$ for $T\ll T_{BG}$, directly reflecting the inelastic nature of low temperature phonon scattering (and the absence of elastic impurity scattering by virtue of purity).  Our results clearly bring this physics out both for 2D and 3D metals.

Our work shows that it is, in principle, possible to use the validity or not of the WF law in order to check the applicability of the FL paradigm through careful measurements with some caveats (and some assumptions about the applicable materials parameter values for the system under consideration).  For example, the quasielastic acoustic phonon scattering for $T>T_{BG}$ invariably produces a temperature dependent electrical resistivity going as linear in $T$.  In a FL, however, this linear-in-$T$ resistivity regime should manifestly obey the WF law as our work shows, provided that $T\ll T_F$ constraint is also satisfied.  So, if a metallic system clearly manifesting a linear-in-$T$ electrical resistivity over a temperature regime also violates the WF law at the same time, this would be a strong indicator of a possible NFL behavior.  Similarly, impurity scattering typically leads to $T$-independent electrical resistivity (again assuming $T_F\gg T$), and therefore, an observed violation of the WF law concomitant with a $T$-independent resistivity (or a linear-in-$T$ resistivity) would be an indicator of a possible NFL behavior.  It may be worthwhile to mention in this context that the cuprate high-$T_c$ superconductors often exhibit a linear-in-T resistivity in the normal phase (although the origin of this linear-in-$T$ resistivity is not agreed upon and is considered by most to be caused by a non-phononic mechanism in contrast to a similar linear-in-$T$ resistivity in normal metals at room temperatures).  The WF law seems to be well-obeyed experimentally by the cuprate systems in the normal phase, indicating that a dominant part of its normal state transport is likely to be of a FL nature, but our lack of understanding of the underlying transport mechanism  makes a definitive conclusion difficult.  It is, however, interesting to note that the cuprates are often referred to as ``strange'' or ``bad'' metals, but the fact that such strange metals seem to obey the WF law is itself rather strange.  One possibility is that the main transport scattering mechanism in the cuprate normal phase arises from spin fluctuations associated with the nearby antiferromagnetic Mott phase, which could provide a simple explanation for the validity of the WF law (as well as the linear-in-$T$ resistivity) assuming that the corresponding Bloch-Gr\"uneisen temperature for the bosonic spin fluctuations is low so that the scattering is primarily quasielastic in nature.  Of course, it is also possible that the linear-in-$T$ resistivity does indeed arise from phonon scattering with a low $T_{BG}$, in which case the WF law emerges naturally.  Obviously, much more work is necessary before a definitive conclusion is possible, and our comments here should be construed only as speculative ideas.

There have been experimental studies of the violation of the WF law in the context of the breakdown of the quasiparticle picture and the FL description.  In most of these studies the WF failure seems to occur near a quantum critical point (e.g. magnetic criticality\cite{tanatar2007anisotropic}, metal-insulator transition \cite{lee2017anomalously}, Dirac point\cite{crossno2016observation}) where the quasiparticle picture may indeed be questionable, but it is also possible that this failure is an inherent effect of electron-electron interactions (neglected in our work) within the Fermi liquid description.  A complete theory of the Wiedemann-Franz behavior leading to a quantitative calculation of $L(T)$ including electron-impurity, electron-phonon, and electron-electron interactions for arbitrary system parameters is a challenging task  which has not been undertaken yet even for a model Fermi liquid, let alone for systems with complicated quantum phase transitions.  Recent work has considered the status of the Wiedemann-Franz law in the presence of electron-electron and electron-impurity interactions (but without any phonons) in continuum Fermi liquids using the hydrodynamic approximation within the Boltzmann theory\cite{LucasPlasmons}.  The key finding is that the ideal WF law is indeed violated at some intermediate temperature range, but the WF law is recovered at low enough temperatures with $L(T)$ going as $L/L_0= \frac{\Gamma}{\gamma + \Gamma}$, where $\Gamma$ ($\gamma$) is respectively the electron-impurity (electron-electron) interaction strength.  Thus, the electron-electron interaction effects vanish in the clean limit ($\Gamma=0$) as it must in the absence of Umklapp and Baber scattering since the electron-electron interaction is manifestly momentum conserving by itself.  This is of course very different from the effect of electron-phonon scattering, where the Lorenz number vanishes at low temperatures in the absence of electron-impurity scattering with $L/L_0 \sim T^2$ for $T\ll T_{BG}$ in a perfectly clean metal.  Using the fact that in a Fermi liquid, $\gamma\sim (T/T_F)^2$ for $T\ll T_F$, we conclude, the violation of the WF law due to the electron-electron interaction is a higher-order effect, going as $L/L_0 \sim (1 - \mathcal{O}(T^2))$ in a dirty system for $T \ll T_F$, whereas the corresponding electron-phonon interaction induced violation of the WF law is a leading-order effect in a clean system, going as $L/L_0 \sim T^2$ for $T\ll T_{BG}$.  This difference arises because the electron-phonon interaction breaks momentum conservation and leads to resistive scattering even without any disorder whereas electron-electron interaction necessitates the presence of disorder  (within the hydrodynamic theory) for breaking the momentum conservation.  (Inclusion of Umklapp scattering in a lattice changes the picture somewhat, but not qualitatively, and is not considered here.)  Thus, in principle it should be possible using  detailed low-temperature ($T\ll T_{BG}$) measurements of $L(T)$ in samples with controlled disorder to distinguish between effects of electron-phonon and electron-electron interactions, but it is likely to be an extremely challenging task.

 Before concluding, we note that inelastic scattering processes considered in the current work always suppress the effective Lorenz number $L(T)$ below the ideal Lorenz number $L_0$, i.e. $L(T)<L_0$ in all our results, a point also emphasized in Ref. \onlinecite{mahajan2013non}.  This implies that inelastic scattering generically enhances the thermal resistivity compared with the electrical resistivity, arising simply from the fact that the electrical resistivity is dominated by large-angle backscattering (``$2k_F$-scattering") across the Fermi surface relaxing the maximum possible momentum whereas the thermal resistivity is affected equally by large-angle and small-angle inelastic scattering processes.  At low temperatures, when $k_B T$ is much smaller than the typical phonon energy, large-angle scattering is strongly suppressed compared with the small-angle scattering, thus enhancing thermal resistivity relative to the electrical resistivity, thus suppressing $L(T)$ well below $L_0$.  This suppression of $L(T)$ well below $L_0$ thus is generic in the presence of strong inelastic scattering independent of the FL or NFL nature of the underlying system.  It is, however, possible for $L(T)$ to exceed $L_0$ in special situations.  Experimentally, this can happen (and often does) when lattice thermal conductivity cannot be separated out from the electronic contribution.  Since the lattice (i.e. phonons themselves) can carry heat rather efficiently, but does not carry any charge, any lattice contribution would enhance the thermal conductivity, making the apparent $L(T)$ exceed $L_0$.  Ensuring that the measured thermal conductivity is all electronic without any lattice contribution whatsoever is a difficult experimental challenge.  Thus, if phonons themselves are conducting heat, the WF law can be violated with the apparent $L(T)>L_0$.  In a similar vein, it is possible for the electrons to lose energy directly to the lattice via electron-phonon interaction through the hot electron energy relaxation process.  Such a direct energy loss from the electrons to the phonons is not a transport or conduction phenomenon, but experimentally this may appear as an enhanced thermal conductivity with $L(T)>L_0$ and an apparent violation of the WF law.  This process could in fact enhance $L(T)$ arbitrarily above $L_0$ unless one is careful.  In the presence of bipolar diffusion (i.e. when both electrons and holes are present in the system in equal numbers), again the thermal conductivity would surpass the WF constraint making $L(T)>L_0$.  In fact, if the electrons and holes are strongly interacting, the $L/L_0$ ratio could be very large as found recently in graphene\cite{crossno2016observation}.  There could be other processes, not considered by us, which could also enhance $L/L_0$ above unity in violation of the WF law.  Our work has focused entirely on the issue of electron-phonon scattering leading to a parametric violation of the WF law at low temperatures in clean systems, where the inelasticity of the scattering process suppresses electrical conductivity much more strongly than the thermal conductivity making $L/L_0 \sim T^2$ at low temperatures in the absence of impurity scattering.

Finally, we  mention several other complications which are likely to cause problems in the study of the WF law in real materials.  In particular, as mentioned above, phonons themselves carry heat (but not electricity) and hence all measurement of $L(T)$ must necessarily ensure that any lattice thermal conductivity contributions are either absent or carefully subtracted out.  This is not an easy task in general.  Second, phonon drag, whence the carriers carry the lattice phonons with them, could be important complicating the extraction of an electronic thermal conductivity.  Similarly, the electrons may not be in equilibrium with the lattice (the so-called ``hot electron effect" mentioned above where the electrons and the phonons are at different temperatures), and in such a situation, the direct energy loss of the electrons to the lattice (through phonon emission for example) may manifest itself as a heat loss from the electrons, but this energy loss (the so-called ‘hot electron energy relaxation’) is completely distinct from the heat diffusion process associated with the electronic thermal conductivity.  It is not always easy to separate hot electron energy loss from electrnic thermal conduction, which may again produce erroneous experimental values of $L(T)$.  Thus, there could be many reasons, some fundamental (e.g. inelastic scattering, nondegeneracy) and some practical (e.g. hot electron energy loss, lattice thermal conductivity), contributing to a breakdown of the WF law, and therefore, it is unwise to automatically accept a breakdown of the WF law (i.e. finding that $L(T)$ differs from $L_0$) as an indicator of an underlying NFL description.  One must carefully consider all the carrier scattering processes contributing to $\kappa$ and $\sigma$ in quantitative depth to see if a FL description with quantitative corrections arising from the details of the scattering processes themselves are leading to the deviation of $L(T)$ from $L_0$.  This is the key message of our work.

\textit{Note added:} A recent work by Hwang and DasSarma \cite{PhysRevB.99.085105}, shows that the linear-in-Tresistivity, often as-sociated with the failure of Fermi liquid paradigm, is alsoconsistent with electron-phonon interactions just as we findthat the breakdown of the WF law may arise from electron-phonon interactions.

\section*{Acknowledgment}
This work is supported by Laboratory for Physical Sciences. A.L. was supported by JQI-PFC-UMD. The authors thank Maissam Barkeshli for several discussions before and during the course of this work.
\bibliography{references}

\appendix
\section{Boltzmann Equation Formalism}\label{app:BEF}

In this appendix, we show the explicit form of the linearized Boltzmann equation, including the collision integral. In all of the appendices we have set $\hbar = k_B = 1$.

In order to use the linearized Boltzmann equation Eq. \eqref{linbol}, the collision integral Eq. \eqref{eqn:collisionIntegral} must also be linearized. Using the detailed balance relation
\begin{equation}
S(\bv{k},\bv{k}')f_0(\bv{k})(1-f_0(\bv{k}')) = S(\bv{k}',\bv{k})(1-f_0(\bv{k}))f_0(\bv{k}')
\end{equation}
the linearized collision integral is
\begin{widetext}
\begin{equation}
\mathcal{I}_{\bv k} = -\int \frac{d^d \bv{k}'}{(2\pi)^d} S(\bv{k},\bv{k}') \left[\delta f(\bv{k}) \frac{1-f_0(\bv{k}')}{1-f_0(\bv{k})} - \delta f(\bv{k}') \frac{f_0(\bv{k})}{f_0(\bv{k}')}\right]
\end{equation}
\end{widetext}
For simplicity, we assume throught this paper an isotropic quadratic band of effective mass $m$. Substituting in the ansatz Eq. \eqref{ansatz}, the linearization allows the integral equation to be broken into separate equations for the thermal and electrical lifetimes:
\begin{widetext}
\begin{align}
1 &= \int \frac{d^d\bv{k}'}{(2\pi)^d} S(\bv{k},\bv{k}') \frac{1-f_0(\varepsilon')}{1-f_0(\varepsilon)}\left[\tau_{\sigma}(\varepsilon) - \frac{k'\cos \alpha'}{k\cos \alpha} \tau_{\sigma}(\varepsilon')\right]
\label{eqn:electricBoltzmannFull}\\
1 &= \int \frac{d^d\bv{k}'}{(2\pi)^d} S(\bv{k},\bv{k}') \frac{1-f_0(\varepsilon')}{1-f_0(\varepsilon)}\left[\tau_{\kappa}(\varepsilon) - \frac{\varepsilon'-\mu}{\varepsilon-\mu}\frac{k'\cos \alpha'}{k\cos \alpha} \tau_{\kappa}(\varepsilon')\right]
\label{eqn:thermalBoltzmannFull}
\end{align}
\end{widetext}
where $\alpha$ (resp. $\alpha'$) is the angle between $\bv{k}$ (resp. $\bv{k}'$) and the applied field.

With a bit of algebra, it is straightforward to show that
\begin{equation}
\frac{\cos \alpha'}{\cos \alpha} = \begin{cases}
\cos \theta + \tan \alpha \sin \theta \cos \phi & d=3\\
\cos \theta - \tan \alpha \sin \theta & d=2
\end{cases}
\end{equation}
where $\theta$ is the angle between $\bv{k}$ and $\bv{k}'$ and $\phi$ is the polar angle for $\bv{k}'$ about $\bv{k}$. In all cases we consider in this paper, $S$ depends only on $k,k',$ and $\cos \theta$, where $\theta$ is the angle between $\bv{k}$ and $\bv{k}'$. Therefore, in $d=3$, the integral over $\phi$ of the $\cos \phi$ term will be zero. In $d=2$, since $S$ depends on $\theta$ only as $\cos \theta$, by orthogonality the $\sin \theta$ term will integrate to zero. The upshot is that we may replace $\cos \alpha'/\cos \alpha$ with $\cos \theta$ in both $d=2$ and $d=3$.

Once these equations have been solved for $\tau$, the transport coefficients may be obtained using Eq. \eqref{tc}.

\section{Impurity Scattering}
\label{app:impurity}

In this appendix we calculate the electrical and thermal conductivities in Boltzmann theory for impurity scattering, including a model where the impurity scattering is purely elastic but also primarily forward. We show that the Wiedemann-Franz law still holds when $T \ll T_F$.

Impurity scattering is elastic, so $S(\bv{k},\bv{k}')\propto \delta(\varepsilon-\varepsilon')$. The lifetime equations Eq. \eqref{eqn:electricBoltzmannFull} and \eqref{eqn:thermalBoltzmannFull} simplify dramatically and actually become the same equation, which is easy to solve:
\begin{equation}
\frac{1}{\tau_{\sigma}(\varepsilon)} = \frac{1}{\tau_{\kappa}(\varepsilon)}= -\int \frac{d^d\bv{k}'}{(2\pi)^d} S(\bv{k},\bv{k}')\left(1-\cos \theta\right)
\label{eqn:elasticBoltzmann}
\end{equation}

We now consider specific impurity scattering models.

\subsection{Conventional Impurity Scattering}

The textbook impurity model is isotropic and short-range with the scattering rate Eq. \eqref{eqn:impurityS}. The Boltzmann equation Eq. \eqref{eqn:elasticBoltzmann} amounts to a simple integral in this case and we find
\begin{equation}
\frac{1}{\tau(\varepsilon)} = \begin{cases}
n_{\text{imp}}mu_0^2 & d=2\\
\frac{n_{\text{imp}} u_0^2 \sqrt{2m^3\varepsilon}}{\pi} & d=3
\end{cases}
\end{equation}
Substitution into Eq. \eqref{tc} leads to the conductivities (per spin)
\begin{widetext}
\begin{align}
\sigma &=e^2A T \log\left(1+e^{\mu/T}\right) \nonumber \\
S &=
\frac{\mu}{T} + \frac{\pi^2}{3\log(1+e^{\mu/T})}	-\log(1+e^{\mu/T})-\frac{2\text{Li}_2((1+e^{\mu/T})^{-1})}{\log(1+e^{\mu/T})} \nonumber \\
L_{TT} &=
A\left[\mu^2\left(\log(1+e^{\mu/T})-4\right)+2T\mu\left(\log^2(1+e^{\mu/T})+\text{Li}_2((1+e^{\mu/T})^{-1})-\frac{2\pi^2}{3}\right)-6T^2\text{Li}_3(-e^{\mu/T})\right]
\label{eqn:impurityConductivities}
\end{align}
\end{widetext}
where the thermopower $S = L_{ET}/\sigma = -L_{TE}/(T\sigma)$ one must remember that the chemical potential $\mu$ is a function of $T$ and
\begin{equation}
A^{-1} = \begin{cases}
2\pi n_{\text{imp}} m u_0^2 & d=2\\
3\pi n_{\text{imp}} m u_0^2 & d=3
\end{cases}
\end{equation}
Note that $n_{\text{imp}}$ and $u_0$ have different units in $d=2$ and $d=3$, and that the functional dependence $\mu(T)$ is different in $d=2$ and $d=3$.

In the regime $T \ll T_F$, $\mu \approx T_F$ and we expand the polylogarithm $\text{Li}_s(-e^x)$ at large values of $x$ using the series representation
\begin{equation}
\text{Li}_s(-e^x) = \sum_{k=0}^{\infty}(-1)^k(1-2^{1-2k})(2\pi)^{2k}\frac{B_{2k}}{(2k)!} \frac{x^{s-2k}}{\Gamma(s+1-2k)} \label{eqn:Sommerfeld}
\end{equation}
where the $B_{2k}$ are the Bernoulli numbers. This series is essentially the Sommerfeld expansion. The resulting conductivities are
\begin{align}
\sigma(T \ll T_F) &= \begin{cases}
\frac{e^2}{4\pi n_{\text{imp}}u_0^2 m^2} & d=2\\
\frac{e^2T_F}{3\pi m n_{\text{imp}}u_0^2} & d=3
\end{cases}\\
\kappa(T \ll T_F) &= \begin{cases}
\frac{\pi T}{12 n_{\text{imp}}u_0^2 m^2} & d=2\\
\frac{\pi T_F}{9 m n_{\text{imp}}u_0^2} & d=3
\end{cases}
\end{align}
and the Wiedemann-Franz law is obeyed ($S$ is of order $T/T_F$ and can be neglected).

At $T \gg T_F$, the temperature dependence of the equilibrium chemical potential $\mu$ must be accounted for. A textbook calculation yields
\begin{align}
\begin{cases}
\mu = T \log\left(e^{E_F/T}-1\right)\approx T \log\left(\frac{E_F}{T}\right) & d=2\\
\mu \approx T\log\left(\frac{4}{3\sqrt{\pi}}\left(\frac{E_F}{T}\right)^{3/2}\right)& d=3
\end{cases}
\end{align}
Plugging this in and expanding Eq. \eqref{eqn:impurityConductivities} leads to
\begin{widetext}
\begin{align}
\sigma &\approx e^2 A \begin{cases}
E_F & d=2\\
\frac{4E_F^{3/2}}{3\sqrt{\pi}}T^{-1/2} & d=3
\end{cases}\\
S &\approx \begin{cases}
2-\log\left(\frac{E_F}{T}\right) & d=2\\
2-\log\left(\frac{4}{3\sqrt{\pi}}\left(\frac{E_F}{T}\right)^{3/2}\right) & d=3
\end{cases}\\
L_{TT} &\approx A \begin{cases}
T E_F \left(6-4\log\left(\frac{E_F}{T}\right)+\log^2\left(\frac{E_F}{T}\right)\right) & d=2\\
\frac{4}{3\sqrt{\pi}}E_F^{3/2}\sqrt{T} \left(6-4\log\left(\frac{4}{3\sqrt{\pi}}\left(\frac{E_F}{T}\right)^{3/2}\right)+\log^2\left(\frac{4}{3\sqrt{\pi}}\left(\frac{E_F}{T}\right)^{3/2}\right)\right) & d=3
\end{cases}
\end{align}
\end{widetext}
The Lorenz number can then be computed straightforwardly for $T \gg T_F$; in both $d=2$ and $d=3$,
\begin{equation}
L = 2 = \frac{6}{\pi^2}L_0
\end{equation}

\subsection{Forward Scattering}
\label{app:forwardImpurity}
We now demonstrate that even when the dominant scattering mechanism is elastic forward scattering (i.e. not isotropic as in Appendix \ref{app:impurity} as above), the Wiedemann-Franz law is still obeyed at low temperature. Thus, pure elastic scattering always leads to the WF law independent of the isotropic or strongly anisotropic nature of the scattering. This result is a special case of what is known on very general grounds from the Sommerfeld expansion\cite{LLKinetics}, but we still find these calculations enlightening; we can show explicitly that even when there is a parameter which we can tune to be in the forward scattering limit, WF is unaffected.

We will use the scattering rate (per unit of momentum space)
\begin{equation}
S(\bv{k},\bv{k}') = U_0^2 n_{\text{imp}} \frac{e^{-2q z_0}}{(q+q_s)^2}\delta(\varepsilon(\bv{k})-\varepsilon(\bv{k}'))
\label{eqn:forwardScatteringRate}
\end{equation}
where $\bv{q} = \bv{k}-\bv{k}'$. Physically, this is the scattering rate obtained from Fermi's Golden Rule for charged impurities placed a distance $z_0$ from a 2D electron gas\cite{AndoRMP}, with $n_{\text{imp}}$ the impurity concentration, $q_s$ a screening wavevector, and $U_0$ a prefactor characterizing the strength of scattering with dimensions of energy times length.  The precise form is unimportant - what matters is that the scattering is elastic and that scattering wavevectors larger than $1/z_0$ are exponentially suppressed. Taking $k_Fz_0 \gg 1$ corresponds to the extreme forward scattering limit\cite{DasSarma1985}.

Substitution into Eq. \eqref{eqn:elasticBoltzmann} yields
\begin{equation}
\frac{1}{\tau(\varepsilon)} = \frac{U_0^2 n_{\text{imp}}}{(2\pi)^d}\int d^d \bv{q} \frac{e^{-2qz_0}}{(q+q_s)^2}\delta\left(\varepsilon(\bv{k})-\varepsilon(\bv{k}+\bv{q})\right)(1-\cos \theta)
\end{equation}
where $\theta$ is the angle between $\bv{k}$ and $\bv{k}'$. We have used $k=k'$ for elastic collisions on a circular Fermi surface to rewrite things in terms of $\theta$.

It is most convenient to use some geometry to find that $1-\cos \theta = -(q/k)\cos \beta$ where $\beta$ is the angle between $\bv{k}$ and $\bv{q}$. Likewise,
\begin{equation}
\varepsilon(\bv{k}) - \varepsilon(\bv{k}+\bv{q}) = -\frac{2kq \cos \beta+q^2}{2m}
\end{equation}

In two dimensions, substituting and changing variables to $u=\cos \beta$ we obtain
\begin{align}
\frac{1}{\tau(\varepsilon)} &= -\frac{U_0^2n_{\text{imp}}}{2\pi^2}\int_0^{\infty}dq \int_{-1}^1 du \frac{u}{\sqrt{1-u^2}} \frac{mq}{k^2} \frac{e^{-2qz_0}}{(q+q_s)^2}\delta\left(u+\frac{q}{2k}\right)\\
&= -\frac{U_0^2mn_{\text{imp}}}{4\pi^2 z_0k^3}\int_0^{2kz_0}dx \frac{x^2}{\sqrt{1-(x/2kz_0)^2}} \frac{e^{-2x}}{(x+q_sz_0)^2} \label{eqn:exactTau}
\end{align}
where we have made the change of variables $x=z_0q$.

Since we are interested in computing the conductivities at $T \ll T_F$, we may take $k \sim k_F$.

In the forward scattering limit $k_Fz_0 \gg 1$, a straightforward series expansion about $x=2kz_0$ shows that the contribution to the integral of the region with $x \sim 2kz_0 \gg 1$ is exponentially suppressed in $k_F z_0$. Therefore, the integral is dominated by the regime $x \ll 2kz_0$. In said regime, the square root factor is, to leading order, 1, so, it is safe to neglect the square root and to extend the upper limit of integration to $+\infty$:
\begin{equation}
\frac{1}{\tau(\varepsilon)} \approx \frac{U_0^2mn_{\text{imp}}}{4\pi^2 z_0k^3}\int_0^{\infty}dx x^2 \frac{e^{-2x}}{(x+q_sz_0)^2} \equiv \frac{1}{A(q_sz_0)\varepsilon^{3/2}} \label{eqn:approxTau}
\end{equation}
The precise form of $A$ is unimportant for the WF law since it is independent of $\varepsilon$ and $T$.

Using Eq. \eqref{tc}, the transport coefficients can be computed explicitly in terms of polylogarithms:
\begin{widetext}
\begin{align}
\sigma &= -\frac{15Ae^2}{16\sqrt{\pi}}T^{5/2}\text{Li}_{5/2}(-e^{T_F/T})\\
\kappa &= -\frac{15A}{64\sqrt{\pi} T}T^{5/2}\left(4T_F^2 \text{Li}_{5/2}(-e^{T_F/T})-28T_F T \text{Li}_{7/2}(-e^{T_F/T})+63T^2\text{Li}_{9/2}(-e^{T_F/T})\right)
\end{align}
\end{widetext}
As we will show in the next subsection,  $T\sigma S^2 \sim (T/T_F)^2 L_{TT}$, so we have taken $\kappa \approx L_{TT}$.

Using the expansion Eq. \eqref{eqn:Sommerfeld},
\begin{align}
\sigma &= \frac{Ae^2}{2\pi}T_F^{5/2}\\
\kappa &= \frac{\pi^2}{3}\frac{A}{2\pi}T_F^{5/2}T
\end{align}
where we used the $k=0$ term for $\sigma$ and the $k=1$ term for $\kappa$ (the $k=0$ term for $\kappa$ is zero, as expected). The Wiedemann-Franz law is obeyed.

\subsection{Forward Scattering: Corrections to WF}

We now want to estimate the leading corrections to the WF law at $T \ll T_F$ in the elastic forward scattering model used in App.~\ref{app:forwardImpurity}. These will be of order $(T/T_F)^2$, arising from doing the next order of the Sommerfeld expansion. Said $(T/T_F)^2$ term will have an order-1 coefficient, but we would also like to obtain the corrections to that coefficient to leading order in $1/k_Fz_0$.

To do so, we need to start by calculating the leading-order corrections to  $\tau$ as a function of $1/k_Fz_0$. As discussed previously, power-law corrections appear only at small $x/2kz_0$ and arise from the lowest-order correction when the square root is expanded. The error is approximately
\begin{equation}
\delta \left(\frac{1}{\tau}\right) \approx -\frac{U_0^2 m n_{\text{imp}}}{4\pi^2 z_0 k^3}\int_0^{\epsilon} dx x^2 \frac{1}{2} \left(\frac{x}{2kz_0}\right)^2 \frac{e^{-2x}}{(x+q_sz_0)^2}
\end{equation}
where $1 \ll \epsilon \ll 2kz_0$ is some cutoff where the expansion of the square root is valid. For the same reasons as before we may take the upper limit to infinity and we obtain
\begin{align}
\delta \left(\frac{1}{\tau}\right) \approx &-\frac{U_0^2 m n_{\text{imp}}}{8\pi^2 z_0 k^3} \frac{1}{(2kz_0)^2}\int_0^{\infty} dx x^4 \frac{e^{-2x}}{(x+q_sz_0)^2} \nonumber \\
&\equiv \frac{1}{B \varepsilon^{5/2}z_0^2}
\end{align}
where we have left implicit the fact that $B$ is a complicated function of $q_s z_0$.

We can now expand
\begin{equation}
\tau = \frac{1}{1/A\varepsilon^{3/2} + 1/B \varepsilon^{5/2}z_0^2} \approx A\varepsilon^{3/2} - \frac{A^2\varepsilon^{1/2}}{B z_0^2}
\end{equation}
where the expansion is controlled by $1/\varepsilon z_0^2 \sim 1/k_F^2 z_0^2$.

This expression can be plugged straightforwardly into Eqs. \eqref{tc}, and we wish to take the next highest order in the Sommerfeld expansion Eq. \eqref{eqn:Sommerfeld}.

We define $\alpha_{3/2} = A$ and $\alpha_{1/2} = A^2/Bz_0^2$. After expanding the polylogarithms to the appropriate order, we find
\begin{align}
\sigma &\approx \frac{e^2}{2\pi}\sum_{n=1/2,3/2} \alpha_n T_F^{n+1}\left(1+\frac{\pi^2}{6}\left(\frac{T}{T_F}\right)^2 n(n+1)\right)\\
L_{ET} &\approx \frac{e}{2\pi} \frac{\pi^2}{3} \left(\frac{T}{T_F}\right) \sum_{n=1/2,3/2} \alpha_n T_F^{n+1}\\
L_{TT} &\approx T \frac{1}{2\pi} \frac{\pi^2}{3} \sum_{n=1/2,3/2} \alpha_n T_F^{n+1} \times \nonumber \\
&\hspace{1.5cm}\left[1-\frac{7\pi^2}{60}\left(\frac{T}{T_F}\right)^2 n^2(n+1)(n+5)\right]
\end{align}
After a considerable amount of algebra and Taylor expansion, we find
\begin{equation}
\frac{L}{L_0} \approx 1 - \left(\frac{T}{T_F}\right)^2 \frac{\pi^2}{24} \left(\frac{1339}{8} + \frac{743}{5}\frac{A}{Bz_0^2 T_F}\right)
\label{eqn:WFCorrection}
\end{equation}
From the definitions,
\begin{equation}
\frac{A}{Bz_0^2 T_F} = \frac{1}{2(k_Fz_0)^2}\frac{I_2(q_sz_0)}{I_4(q_sz_0)}
\end{equation}
with
\begin{equation}
I_n(q_sz_0) = \int_0^{\infty} x^n \frac{e^{-2x}}{(x+q_sz_0)^2}
\end{equation}
We note that the correction to the WF law arising in Eq.~\eqref{eqn:WFCorrection} from the forward scattering physics is of $\mathcal{O}(T/T_F)^2$, which is the same order where electron-electron scattering also shows up as a correction\cite{lucas2018electronic} of the WF law, thus considerably complicating interpretation of experiments.

\section{Electron-Phonon Transport Calculations}\label{app:phonon}

In this appendix, we discuss our Boltzmann theory electron-phonon calculations in detail. Throughout we assume a quadratic band of effective mass $m$ and the scattering rate
\begin{equation}
S(\bv{k},\bv{k}') = \frac{\pi D^2 q^2}{\rho \omega_q}\left[N_q \delta(\epsilon - \epsilon' + \omega_q) + (N_q+1)\delta(\epsilon-\epsilon'-\omega_q)\right]\Theta(\omega_D - \omega_q)
\end{equation}
obtained by Fermi's Golden Rule for electrons of momentum $\bv{k}$ scattering off of acoustic phonons. Here $\bv{q}$ is the momentum transfer, equal to $\bv{k}'-\bv{k}$ in the first term and equal to $\bv{k}-\bv{k}'$ in the second term. Also, $N_q$ is the Bose distribution, $D$ is the deformation potential, $\omega_q = c_s q$, $c_s$ is the speed of sound in the material, $\omega_D$ is the Debye frequency, and $\Theta$ is the Heaviside step function. We assume throughout that the system is sufficiently clean so that electron-impurity scattering can be neglected at the temperatures in question. We also neglect effects such as phonon drag. (Note also that for the results in our main text we assume $T_{BG}<T_D$ throughout so that the effective phonon frequency cut off is $T_{BG}$ for our analysis.)

\subsection{Relaxation Time Approximations}

In principle, the integral equation Eq. \eqref{eqn:electricBoltzmannFull} can be solved. As we have seen from Appendix \ref{app:impurity}, this is straightforward when the scattering is purely elastic. However, electron-phonon scattering is inelastic, so the Boltzmann equation remains a complicated integral equation for $\tau$. To make progress, we need to perform an uncontrolled approximation on Eqs. \eqref{eqn:electricBoltzmannFull} and \eqref{eqn:thermalBoltzmannFull}. In particular, we will replace $\tau_{\sigma,\kappa}(\varepsilon') \rightarrow \tau_{\sigma,\kappa}(\varepsilon)$. Although the terminology is used in ambiguous or inconsistent ways in the literature, this is our form of the ``relaxation time approximation."

With this approximation the Boltzmann equation becomes
\begin{align}
\frac{1}{\tau_{\sigma}(\varepsilon)} &= \int \frac{d^d\bv{k}'}{(2\pi)^d} S(\bv{k},\bv{k}') \frac{1-f_0(\varepsilon')}{1-f_0(\varepsilon)}\left[1 - \frac{k'}{k} \cos \theta \right] \nonumber \\
\frac{1}{\tau_{\kappa}(\varepsilon)} &= \int \frac{d^d\bv{k}'}{(2\pi)^d} S(\bv{k},\bv{k}') \frac{1-f_0(\varepsilon')}{1-f_0(\varepsilon)}\left[1 - \frac{\varepsilon'-\mu}{\varepsilon-\mu}\frac{k'}{k} \cos \theta \right]
\label{eqn:rta}
\end{align}

Plugging in the form of $S(\bv{k},\bv{k}')$ and using the expressions
\begin{align}
\epsilon-\epsilon' &= -\frac{q^2\pm 2kq \cos \beta}{2m}\\
1-\frac{k'}{k}\cos \theta &= \mp \frac{q}{k} \cos \beta
\end{align}
where the sign corresponds with $\bv{k}' = \bv{k}\pm \bv{q}$ (depending on whether a phonon is being absorbed or emitted), we find
\begin{widetext}
\begin{align}
\frac{1}{\tau_{\sigma}(\varepsilon)} = \frac{\pi D^2 m}{\rho c_s k}\int \frac{d^d\bv{q}}{(2\pi)^d} \left[\frac{1-f_0(\varepsilon+ c_s q)}{1-f_0(\varepsilon)}N_q\left(-\frac{q}{k}\cos \beta\right)\delta\left(\cos \beta - \frac{-q+2m c_s}{2k}\right) +\right.\nonumber\\
\left. + \frac{1-f_0(\varepsilon- c_s q)}{1-f_0(\varepsilon)}(N_q+1)\left(  \frac{q}{k}\cos \beta\right)\delta\left(\cos \beta - \frac{q+2m c_s}{2k}\right)\right]\Theta(q_D - q)
\label{eqn:generalElectricRTA}
\end{align}
\end{widetext}
where $q_D$ is the Debye wavevector $\omega_D/c_s$. Similar substitutions can be made for the thermal lifetime.

We discuss these approximations further in Appendix \ref{app:approximations}

\subsection{3D Calculations}

We start with the electrical conductivity. The angular integrals in Eq. \eqref{eqn:generalElectricRTA} are done mostly straightforwardly, with one important caveat. Since $\cos \beta$ is only integrated over the range $(-1,1)$, the delta functions only lead to nonzero contributions for certain values of $q$; this restriction is where the Bloch-Gruneisen temperature plays a key role. We find
\begin{widetext}
\begin{align}
\frac{1}{\tau(\varepsilon)} = \frac{D^2 m}{4\pi\rho c_s k} \left[\int_{q_{min}^{(1)}}^{q_{max}^{(1)}} dq q^2 \frac{1-f_0(\varepsilon+ c_s q)}{1-f_0(\varepsilon)}N_q\left( -\frac{mc_s q}{k^2} + \frac{q^2}{2k^2}\right) + \int_0^{q_{max}^{(2)}} dq q^2
 \frac{1-f_0(\varepsilon- c_s q)}{1-f_0(\varepsilon)}(N_q+1)\left(\frac{mc_s q}{k^2} + \frac{q^2}{2k^2}\right)\right]
\end{align}
\end{widetext}
Under the assumption $T_D \gg T_{BG}$ (and noting that in most systems $T_{BG} \ll T_F$),
\begin{align}
q_{min}^{(1)} &= -2k + 2mc_s\\
q_{max}^{(1)} &= 2k+2mc_s\\
q_{max}^{(2)} &= 2k-2mc_s
\end{align}
Defining $\eta = (\varepsilon-\mu)/T$ and $z=c_s q/T$,
\begin{widetext}
\begin{align}
\frac{1}{\tau_{\sigma}(\varepsilon)} = \frac{D^2 mT^3}{4\pi \rho c_s^4 k}\left[\int_{z_{min}^{(1)}}^{z_{max}^{(1)}} dz \frac{1+e^{\eta}}{1+e^{\eta+z}}\frac{z^2}{1-e^{-z}}\left(-\frac{m T z}{k^2} + \frac{T^2 z^2}{2k^2c_s^2}\right)  + \int_0^{z_{max}^{(2)}} dz
 \frac{1+e^{\eta}}{1+e^{\eta-z}}\frac{z^2}{e^z-1}\left(\frac{m T z}{k^2} + \frac{T^2 z^2}{2k^2c_s^2}\right)\right]
 \label{eqn:zIntegral3D}
\end{align}
\end{widetext}
with the definitions of $z_{max,min}$ following from those of $q_{max,min}$.

A very similar computation for the thermal transport lifetime yields
\begin{widetext}
\begin{align}
\frac{1}{\tau_{\kappa}(\varepsilon)} = \frac{D^2 mT^3}{4\pi \rho c_s^4 k}\left[\int_{z_{min}^{(1)}}^{z_{max}^{(1)}} dz \frac{1+e^{\eta}}{1+e^{\eta+z}}\frac{z^2}{1-e^{-z}}\left(1-\frac{\eta + z}{\eta}\left(1+\frac{m T z}{k^2} - \frac{T^2 z^2}{2k^2c_s^2}\right)\right) +\right. \nonumber\\
\left. + \int_0^{z_{max}^{(2)}} dz
 \frac{1+e^{\eta}}{1+e^{\eta-z}}\frac{z^2}{e^z-1}\left(1-\frac{\eta - z}{\eta}\left(1-\frac{m T z}{k^2} - \frac{T^2 z^2}{2k^2c_s^2}\right)\right)\right] \label{eqn:zGammaIntegral3D}
\end{align}
\end{widetext}
To make progress, we now need to look at asymptotic regimes.

\subsubsection{Equipartition Regime in 3D}

This regime is the traditional $T$-linear resistivity regime: $T_{BG} \ll T \ll T_F$. When $T \ll T_F$, only $\varepsilon \sim T_F$ is important so we may estimate $k \sim k_F$. Then to leading order $z_{max}^{(1)} \approx z_{max}^{(2)} \approx 2k c_s/T \sim T_{BG}/T \ll 1$, and $z_{min}^{(1)} = 0$. Since $z \in (0,z_{max})$ and $z_{max} \ll 1$, we may expand Eq. \eqref{eqn:zIntegral3D} to the lowest nontrivial order in $z$:
\begin{align}
\frac{1}{\tau_{\sigma}(\varepsilon)} &\approx \frac{D^2 mT^3}{4\pi \rho c_s^4 k}\int_0^{z_{max}} dz \frac{z^3 T^2}{k^2c_s^2}\left(1+\frac{mc_s^2}{T}\tanh(\eta/2)\right) \\
&\approx \frac{D^2 m Tk}{\pi \rho c_s^2}
\end{align}
where we neglected the term of order $mc_s^2/T \sim T_{BG}^2/TT_F \ll 1$.

The conductivity is computed straightforwardly in the lowest-order Sommerfeld expansion to obey a Drude formula
\begin{equation}
\sigma = \frac{ne^2 \tau(\varepsilon_F)}{m} = \frac{8e^2\rho k_F^2 c_s^2}{6\pi D^2 m^2 T}
\end{equation}
This is the familiar result that the electron-phonon scattering induced resistivity goes as $T$ at high temperatures where the phonons are in the equipartition regime.  Strictly speaking, this linear-in-$T$ regime applies for $T \gtrsim T_{BG}/5$ (or $T_D/5$) depending on whether $T_{BG}<T_D$ or not.

In calculating the thermal lifetime, we can similarly equate the limits of the two integrals in Eq. \eqref{eqn:zGammaIntegral3D} and expand. We find
\begin{widetext}
\begin{align}
\frac{1}{\tau_{\kappa}(\varepsilon)} &\approx \frac{D^2 m T^3}{4\pi \rho c_s^4 k}\int_0^{z_{max}} dz \frac{z^3 T^2}{k^2 c_s^2}\left(1+\frac{mc_s^2}{T}\tanh(\eta/2)-\frac{2c_s^2 m}{T\eta}+\frac{c_s^2	k^2}{T^2 \eta}\tanh(\eta/2)\right)
\end{align}
\end{widetext}
In calculating the thermal conductivity, $\tau_{\kappa}$ is integrated against $(\varepsilon-\mu)^2 \partial f_0/\partial \varepsilon$, which is peaked at $\varepsilon-\mu \sim T$ and equal to zero at $\varepsilon = \mu$. Therefore, when calculating $\tau_{\kappa}$, we can safely estimate $\varepsilon-\mu \sim T$, that is, $\eta \sim 1$, when estimating which terms are important (as long as $\tau_{\kappa}$ does not diverge at $\varepsilon \rightarrow \mu$).

For $\eta \sim 1$ we can neglect all of the $\eta$-dependent terms, which are of order $T_{BG}^2/TT_F \ll 1$ or $T_{BG}^2/T^2 \ll 1$. We find $\tau_{\kappa} \approx \tau_{\sigma}$, in agreement with our results in the main text. The Sommerfeld expansion immediately leads to the Wiedemann-Franz law. Thus, a linear-in-$T$ resistivity arising from phonon scattering is automatically associated with the validity of the WF law.

\subsubsection{Bloch-Gruneisen Regime in 3D}

This regime is $T \ll T_{BG} \ll T_F$. As before, $z_{max}^{(1)} \approx z_{max}^{(2)} \approx 2k c_s/T$, but in this regime both of these limits are large. Since the integrand in Eq. \eqref{eqn:electricBoltzmannFull} is suppressed exponentially at large $z$, it is a good approximation to take $z_{max} \rightarrow \infty$.

After taking $z \rightarrow -z$ in the second term of Eq. \eqref{eqn:electricBoltzmannFull} we obtain
\begin{widetext}
\begin{align}
\frac{1}{\tau_{\sigma}(\varepsilon)} &= \frac{D^2 mT^3}{4\pi \rho c_s^4 k}\int_{-\infty}^{\infty} dz \frac{1+e^{\eta}}{1+e^{\eta+z}}\frac{z^2}{|1-e^{-z}|}\left(-\frac{m T z}{k^2} + \frac{T^2 z^2}{2k^2c_s^2}\right)\\
&= \frac{3D^2 m T^5}{4\pi \rho c_s^6 k^3} \left[\frac{2mc_s^2}{T}\left(\text{Li}_4(-e^{-\eta})-\text{Li}_4(-e^{\eta})\right)+4\left(2\zeta(5)-\text{Li}_5(-e^{-\eta})-\text{Li}_5(-e^{\eta})\right)\right]
\end{align}
\end{widetext}

Again, the electrical conductivity is found at leading order in the Sommerfeld expansion
\begin{equation}
\sigma = \frac{ne^2 \tau(\varepsilon_F)}{m} = \frac{2\rho c_s^6 k_F^6}{9\pi m^2 D^2}\frac{1}{T^5} = \frac{\rho T_{BG}^6}{288 m^2 D^2}\frac{1}{T^5}
\end{equation}
which leads to the expected $\rho \sim T^5$ behavior (often called the Bloch-Gruneisen behavior).

The same approximations can be used in calculating the thermal lifetime
\begin{widetext}
\begin{align}
\frac{1}{\tau_{\kappa}(\varepsilon)} &= \frac{D^2 m T^3}{4\pi \rho c_s^4 k}\int_{-\infty}^{\infty} dz \frac{1+e^{\eta}}{1+e^{\eta+z}}\frac{z^2}{|1-e^{-z}|}\left[1-\left(1+\frac{z}{\eta}\right)\left(1+\frac{mTz}{k^2}-\frac{T^2 z^2}{2k^2c_s^2}\right)\right]
\end{align}
\end{widetext}
In the $z \lesssim 1$ regime where the integrand is not exponentially suppressed, we can use $mT/k^2 \sim T/T_F \ll 1$ and $T^2/k^2c_s^2 \sim T^2/T_{BG}^2 \ll 1$ to simplify the integral dramatically for $\eta \sim 1$:
\begin{align}
\frac{1}{\tau_{\kappa}(\varepsilon)} &\approx \frac{D^2 m T^3}{4\pi \rho c_s^4 k}\int_{-\infty}^{\infty} dz \frac{1+e^{\eta}}{1+e^{\eta+z}}\frac{z^2}{|1-e^{-z}|}\left(-\frac{z}{\eta}\right)\\
&= \frac{6D^2 m T^3}{4\pi \rho c_s^4 k \eta}\left(\text{Li}_4(-e^{-\eta})-\text{Li}_4(-e^{\eta})\right)
\end{align}
In the Sommerfeld expansion, the leading-order contribution is zero as expected. The next-leading-order contribution yields
\begin{equation}
\kappa \approx L_{TT} \approx \frac{8\pi^2\rho c_s^4T_F^2}{54\zeta(3)D^2} \frac{1}{T^2}
\end{equation}
The numerical prefactor should, of course, not be taken very seriously, but we obtain the $1/T^2$ behavior as expected. The Wiedemann-Franz law is violated as
\begin{equation}
L \sim T^2
\end{equation}
The scalings $\sigma \sim 1/T^5$, $\kappa \sim 1/T^2$, and $L \sim T^2$ are in agreement with the calculations in the main text. Thus, in the Bloch-Gruneisen regime the WF law is violated strongly as long as impurity scattering contribution to resistivity is much smaller than the phonon scattering contribution-- in other words, any observation of a Bloch-Gruneisen transport behavior must automatically be associated with a strong violation of the WF law.

\subsection{2D Calculations}

The angular integral in Eq. \eqref{eqn:generalElectricRTA} is slightly more tedious in 2D. Changing variables to $u=\cos \beta$ introduces a factor of 2 and a Jacobian. In the same variables $z=c_s q/T$ and $\eta = (\varepsilon-\mu)/T$ as for 3D, we obtain
\begin{widetext}
\begin{align}
\frac{1}{\tau_{\sigma}(\epsilon)} = \frac{D^2m T^2}{\pi \rho c_s^3 k}&\left[\int_{z_{min}^{(1)}}^{z_{max}^{(1)}} dz \frac{z}{(1-e^{-z})\sqrt{1-\left(\frac{T}{2kc_s}\right)^2\left(z-\frac{2mc_s^2}{T}\right)^2}}\frac{1+e^{\eta}}{1+e^{\eta+z}}\left(-\frac{mTz}{k^2}+\frac{T^2z^2}{2k^2c_s^2}\right) \right. \nonumber \\
&\left.+\int_{0}^{z_{max}^{(2)}} dz \frac{z}{(e^{z}-1)\sqrt{1-\left(\frac{T}{2kc_s}\right)^2\left(-z-\frac{2mc_s^2}{T}\right)^2}}\frac{1+e^{\eta}}{1+e^{\eta-z}}\left(\frac{mTz}{k^2}+\frac{T^2z^2}{2k^2c_s^2}\right)\right] \label{eqn:2DFullIntegral}\\
\frac{1}{\tau_{\kappa}(\epsilon)} = \frac{D^2m T^2}{\pi \rho c_s^3 k}&\left[\int_{z_{min}^{(1)}}^{z_{max}^{(1)}} dz \frac{z}{(1-e^{-z})\sqrt{1-\left(\frac{T}{2kc_s}\right)^2\left(z-\frac{2mc_s^2}{T}\right)^2}}\frac{1+e^{\eta}}{1+e^{\eta+z}}\left(1-\frac{\eta+z}{\eta}\left(1+\frac{mTz}{k^2}-\frac{T^2z^2}{2k^2c_s^2}\right)\right) \right. \nonumber \\
&\left.+\int_{0}^{z_{max}^{(2)}} dz \frac{z}{(e^{z}-1)\sqrt{1-\left(\frac{T}{2kc_s}\right)^2\left(-z-\frac{2mc_s^2}{T}\right)^2}}\frac{1+e^{\eta}}{1+e^{\eta-z}}\left(1-\frac{\eta-z}{\eta}\left(1-\frac{mTz}{k^2}-\frac{T^2z^2}{2k^2c_s^2}\right)\right)\right] \label{eqn:2DFullGammaIntegral}
\end{align}
\end{widetext}
with the limits of the integrals defined in the same way as in 3D.

We must take limits carefully to proceed.

\subsubsection{Equipartition Regime in 2D}
As in 3D, this regime is $T_{BG} \ll T \ll T_F$, which has $z_{max}^{(1)} \sim z_{max}^{(2)} \sim T_{BG}/T \ll 1$ and $z_{min}^{(1)} = 0$. However, the square root makes the integrals a bit more complicated. We expand only the exponentials in $z$ to obtain
\begin{widetext}
\begin{align}
\frac{1}{\tau_{\sigma}(\epsilon)} &\approx \frac{D^2m T^4}{2\pi \rho c_s^5 k^3}\left[\int_{0}^{z_{max}^{(1)}} dz \frac{z}{\sqrt{1-\left(\frac{T}{2kc_s}\right)^2\left(z-\frac{2mc_s^2}{T}\right)^2}}\left(z-\frac{2mc_s^2}{T}\right) \right. \nonumber \\
&\left.\hspace{1in}+\int_{0}^{z_{max}^{(2)}} dz \frac{z}{\sqrt{1-\left(\frac{T}{2kc_s}\right)^2\left(-z-\frac{2mc_s^2}{T}\right)^2}}\left(z+\frac{2mc_s^2}{T}\right)\right]\\
&=\frac{D^2m T^4}{2\pi \rho c_s^5 k^3}\sum_{\pm}\int_{\mp 2mc_s^2/T}^{2kc_s/T} dz \frac{z\left(z\pm \frac{2mc_s^2}{T}\right)}{\sqrt{1-\left(\frac{zT}{2kc_s}\right)^2}}\\
&\approx \frac{2 D^2 m T}{c_s^2 \rho}
\end{align}
\end{widetext}
where we changed variables $z \rightarrow z \pm 2mc_s^2/T$.
Again we can use the Sommerfeld expansion to lowest order to obtain a Drude-type formula
\begin{equation}
\sigma = \frac{ne^2 \tau(\varepsilon_F)}{m} = \frac{e^2 \rho c_s^2k_F^2}{8\pi D^2 m^2}\frac{1}{T}
\end{equation}
with the electrical resistivity linear in temperature.

Doing a similar expansion for $\tau_{\kappa}$ we obtain
\begin{widetext}
\begin{align}
\frac{1}{\tau_{\kappa}(\epsilon)} \approx \frac{D^2m T^2}{\pi \rho c_s^3 k}&\left[\int_{z_{min}^{(1)}}^{z_{max}^{(1)}} dz \frac{1}{\sqrt{1-\left(\frac{T}{2kc_s}\right)^2\left(z-\frac{2mc_s^2}{T}\right)^2}}\left(1-\frac{\eta+z}{\eta}\left(1+\frac{mTz}{k^2}-\frac{T^2z^2}{2k^2c_s^2}\right)\right) \right. \nonumber \\
&\left.+\int_{0}^{z_{max}^{(2)}} dz \frac{1}{\sqrt{1-\left(\frac{T}{2kc_s}\right)^2\left(-z-\frac{2mc_s^2}{T}\right)^2}}\left(1-\frac{\eta-z}{\eta}\left(1-\frac{mTz}{k^2}-\frac{T^2z^2}{2k^2c_s^2}\right)\right)\right]
\approx \frac{2D^2mT}{c_s^2\rho} = \frac{1}{\tau_{\sigma}(\varepsilon)}
\end{align}
\end{widetext}
where we have neglected terms of order $T_{BG}^2/(TT_F) \ll 1$ (and again taken $\eta \sim 1$ when estimating the size of terms). Since the thermal and electrical lifetimes are equal and energy-independent, in agreement with the results in the main text, the Sommerfeld expansion immediately tells us that the Wiedemann-Franz law is obeyed.

\subsubsection{Bloch-Gruneisen Regime in 2D}
As in 3D, this regime is $T \ll T_{BG} \ll T_F$. We again have $z_{min}^{(1)} = 0 $ and $z_{max}^{(1)} \approx z_{max}^{(2)} \sim T_{BG}/T$, but now $z_{max} \gg 1$. If, as in 3D, we wish to take $z_{max} \rightarrow \infty$, we must deal carefully with the square root factor.

When $z \sim z_{max} \gg 1$, the whole integrand is exponentially suppressed, although there is a divergent prefactor scaling as $(z-z_{max})^{-2}$. It can be checked in a straightforward Taylor expansion that contribution of the large-$z$ regime is finite and exponentially small in $z_{max}$. At $z \ll z_{max}$, the term under the square root is of order $1-(z/(T_{BG}/T) \pm T_{BG}/T_F)^2$. The correction to 1 is small whenever $z \ll z_{max}$ since $z_{max} \leq T_{BG}/T$. Therefore, over the entire region of integration, the square root may simply be set to 1. Note that this argument holds for both the electrical and thermal relaxation times.

It is then safe to take $z_{max} \rightarrow \infty$. With a substitution $z \rightarrow -z$ in the second integral, Eq. \eqref{eqn:2DFullIntegral} becomes
\begin{widetext}
\begin{align}
\frac{1}{\tau_{\sigma}(\epsilon)} &= \frac{D^2m T^4}{2\pi \rho c_s^5 k^3} \int_{-\infty}^{\infty} dz \frac{z^2}{1-e^{-z}}\frac{1+e^{\eta}}{1+e^{\eta+z}}\left(z-\frac{2mc_s^2}{T}\right)\\
&=\frac{D^2m T^4}{2\pi \rho c_s^5 k^3} \left[ \frac{2\pi^4}{15}-6\text{Li}_4(-e^{\eta})-6\text{Li}_4(-e^{-\eta})-\frac{4mc_s^2}{T}\left(\text{Li}_3(-e^{\eta})-\text{Li}_3(-e^{-\eta}\right)\right]
\end{align}
\end{widetext}
The electrical conductivity is again found in the lowest-order term of the Sommerfeld expansion
\begin{equation}
\sigma = \frac{ne^2 \tau(\varepsilon_F)}{m} = \frac{2e^2 \rho c_s^5 k_F^5}{\pi^4 D^2 m^2 T^4}
\end{equation}
leading to $\rho \sim T^4$ as expected.

The thermal lifetime, under the same approximations, is
\begin{widetext}
\begin{equation}
\frac{1}{\tau_{\kappa}(\varepsilon)} \approx \frac{D^2mT^2}{\pi \rho c_s^3 k}\int_{-\infty}^{\infty} \frac{z}{1-e^{-z}}\frac{1+e^{\eta}}{1+e^{\eta+z}} \left(1-\frac{\eta+z}{\eta}\left(1+\frac{mTz}{k^2}-\frac{T^2z^2}{2k^2c_s^2}\right)\right)
\end{equation}
\end{widetext}
In the regime $|z| \lesssim 1$ and with $\eta \sim 1$, the leading-order term in the parentheses is $-z/\eta$. Since the integrand is exponentially suppressed at $z \gtrsim 1$,
\begin{align}
\frac{1}{\tau_{\kappa}(\varepsilon)} &\approx \frac{D^2mT^2}{\pi \rho c_s^3 k}\int_{-\infty}^{\infty} \frac{z}{1-e^{-z}}\frac{1+e^{\eta}}{1+e^{\eta+z}} \left(-\frac{z}{\eta}\right)\\
&=\frac{2D^2 m T^2}{\pi \rho c_s^3 k \eta}\left(\text{Li}_3(-e^{-\eta})-\text{Li}_3(-e^{\eta})\right)
\end{align}
The Sommerfeld expansion yields
\begin{equation}
\kappa \approx L_{TT} \approx \frac{c_s^3 T_F^{3/2} \rho}{\sqrt{2}D^2 \sqrt{m}} \frac{1}{T}
\end{equation}
which leads to a violation of the Wiedemann-Franz law
\begin{equation}
L \sim T^2
\end{equation}
Again, the scalings $\sigma \sim 1/T^4$, $\kappa \sim 1/T$, and $L \sim T^2$ all agree with the results in the main text. Thus, both in 2D and 3D FL systems, the WF law is obeyed (violated) in the linear-in-$T$ high-temperature (Bloch-Gruneisen low temperature) regime as long as impurity scattering remains weak.

\section{WF law violation with single relaxation time}
\label{app:wf_single}
It was shown in the main text that thermal relaxation time can differ significantly from charge relaxation time due to different  mechanisms underlying each relaxation process which results in WF law violation. In this appendix we will show that even with a single relaxation time, WF law could still be violated at very low temperatures.

Derivation of WF law  at low temperatures in systems which are described by a single relaxation time, relies on the Sommerfeld expansion of the listed integrals in Eq.\eqref{tc}\cite{ashcroft2005solid}. Generally, the Sommerfeld expansion can be used to evaluate low temperature limits of any integral which involves Fermi distribution function:
\begin{align}
  \int_{-\infty}^{+\infty}\frac{H(\varepsilon)}{e^{(\varepsilon-\mu)/T}+1}\dd \varepsilon=&\int_{-\infty}^{\mu}H(\varepsilon)\dd \varepsilon+\frac{\pi^2}{6}T^2\, H'(\mu)\nonumber \\&+\mathcal{O}\qty(\frac{T}{\mu})^4
\end{align}

For the expansion to be controlled by $\qty(\frac{T}{\mu})$, one needs to make sure that derivatives of $H(\varepsilon)$ do not involve powers of $\frac{1}{T}$. However, non-trivial energy dependences in $H(\varepsilon)$ could result in such factors. For example, terms like $e^{(\varepsilon-\mu)/T}$ in $H(\varepsilon)$ could potentially makes keeping first few terms in Sommerfeld expansion incorrect. As we will show bellow, this could be the case whenever relaxation time involves exponential factors related to statistical distributions.

Consider the system studied in section \ref{sec:aph} and Appendix \ref{app:phonon}, where only electron-phonon scattering is present. To arrive at the integral expressions in Eq. \eqref{eqn:rta} using the RTA, we replaced $\tau_{\sigma,\kappa}(\varepsilon')$ in Eqs. \eqref{eqn:electricBoltzmannFull} and \eqref{eqn:thermalBoltzmannFull} by $\tau_{\sigma,\kappa}(\varepsilon)$. However, if we use a different type of RTA and replace $(\varepsilon'-\mu)\tau_{\kappa}(\varepsilon')\to(\varepsilon-\mu)\tau_{\kappa}(\varepsilon)$
and $\tau_{\sigma}(\varepsilon')\to\tau_{\sigma}(\varepsilon)$, we will find that the two relaxation times become equal and both can be evaluated from the expression for $\tau_\sigma^{-1}(\varepsilon)$ in Eq. \eqref{eqn:rta}.

\begin{figure*}
  \includegraphics[width=\textwidth]{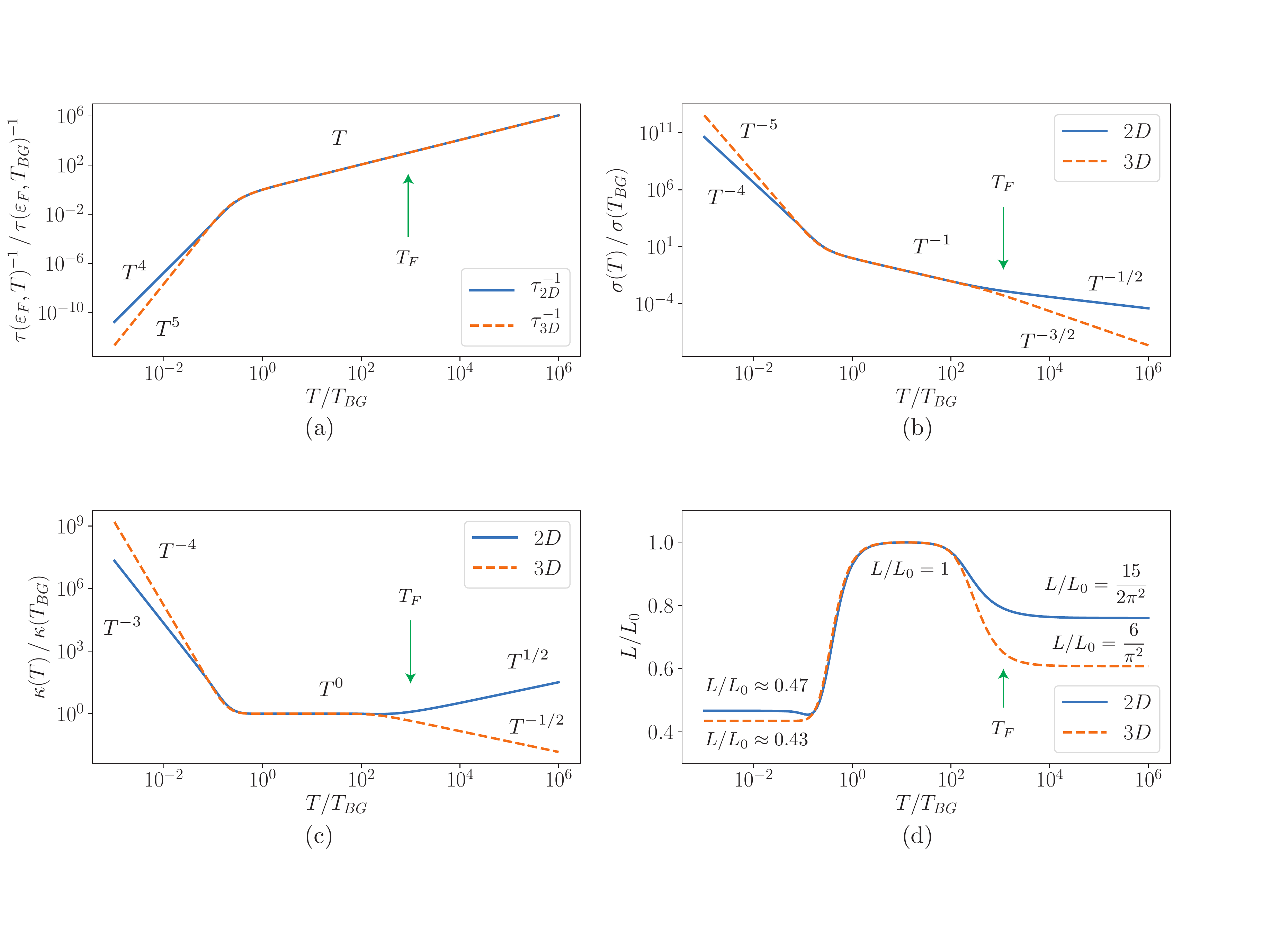}
  \caption{Transport properties of a clean, coupled electron-phonon system using a single-relaxation-time version of the relaxation time approximation. (a) Scattering rate $\tau^{-1}$ at the Fermi energy versus temperature. (b) Electrical conductivity $\sigma$ versus temperature. (c) Thermal conductivity $\kappa$ versus temperature. (d) Lorenz ratio $L/L_0$ versus temperature. In all graphs, solid and dashed lines correspond to 2D and 3D systems respectively. Relaxation rates and transport coefficients are plotted relative to their values at $T=T_{BG}$ to make them dimensionless. The Fermi energy is chosen such that $T_F=10^3 T_{BG}$}
  \label{fig:aph-single}
\end{figure*}

Using this single relaxation time, transport coefficients and hence the Lorenz ratio can be evaluated easily using Eq. \eqref{tc}. Figure \ref{fig:aph-single} shows the result for both 2D and 3D systems.

The plot in Fig.\ref{fig:aph-single}a is exactly the same as the plots for $\tau_\sigma$ in Fig.\ref{fig:aph}a as one would expect. However this time we have only a single relaxation time characterizing both transports. Fig.\ref{fig:aph-single}b is also identical to Fig.\ref{fig:aph}b. The thermal conductivity for $T \gg T_{BG}$ remains the same, but for $T \ll T_{BG}$ we get a different temperature scaling such that the Lorenz ratio $L=\kappa/(\sigma T)$ becomes independent of temperature. However, as one can see from Fig.\ref{fig:aph-single}d, $L/L_0$ no longer saturates to unity but rather approaches a value which is almost half of what WF law predicts. This number can be expressed in terms of definite integrals over polylogarithm functions and turns out to be:
\begin{equation}
    \frac{L}{L_0}\approx\begin{cases}
      0.47\quad &\text{2D}\\
      0.43  &\text{3D}
    \end{cases},\qquad (T \ll T_{BG})
\end{equation}
Generally the exact value depends on the specific form of interactions but is independent of system parameters. Note that the $T \ll T_{BG}$ regime is exactly where the exponential factors in phonons' distribution function become important, which in turn makes the Sommerfeld expansion inapplicable. It is worth noting that modification of the Lorenz number due to the energy dependence of relaxation times has already been discussed in the context of electron-electron scattering\cite{pourovskii2017electron, herring1967simple}.

Regardless of the validity of the RTA which is used in this section, the main point is that the WF law could still be violated while both thermal and electrical transports are described by a single relaxation time. The validity of the RTA in general is discussed in Appendix \ref{app:approximations}.

\section{Discussion of the Approximations}
\label{app:approximations}

The relaxation time approximation (RTA), as implemented in Eqs.~\eqref{eqn:rta} is uncontrolled. One could easily imagine repeating the calculations with a slightly different ansatz (for example, absorbing the factor of $(\varepsilon-\mu)/T$ into $\tau_{\kappa}$ in Eq. \eqref{ansatz}); doing so can in fact lead to qualitatively different results. As such, we should give some justification for our choices. Note that in general the Boltzmann equation, being an integral equation, can be numerically solved iteratively, but such an iterative numerical solution has no mathematical transparency or physical understanding, forcing us to resort to the RTA which provides qualitatively correct, but numerically inaccurate, results.

First, we note that, as discussed in Appendix~\ref{app:impurity}, the RTA in the form we have used is exact when the scattering is elastic and isotropic. Therefore, in the high-temperature equipartition regime $T \gg T_{BG}$ where the scattering is quasi-elastic, the form of the RTA we have used is physically justified. Furthermore, in the low-temperature regime $T \ll T_i \ll T_{BG}$, a controlled perturbative calculation is available~\cite{wilson1954theory}. In this regime, Matthiessen's rule is approximately valid, so the phonon contribution to the transport coefficients can be disentangled from the impurity contribution. The phonon contribution found in the perturbative approach is in qualitative agreement with our RTA results, both from the main text and Appendix~\ref{app:phonon}. Our choice of RTA is a good one in the sense that choosing other forms of the RTA will often lead to qualitative \textit{disagreement} with the perturbative calculation in the regime where it is valid; see Appendix \ref{app:wf_single} for a one-relaxation-time example.

Our choice of RTA therefore yields qualitative agreement with controlled results in the high- and low-temperature regimes. We therefore expect that our RTA results should give qualitatively correct results when interpolating between these two regimes, in particular in the regime of interest $T_i \ll T \ll T_{BG}$ (except if there are regimes where other energy scales become important). As we are not concerned with quantitative predictions, this is sufficient for our purposes: to show that parametrically large violations of the Wiedemann-Franz law can occur in ordinary metals in a regime set by an energy scale $T_{BG}$ which may differ dramatically from $T_D$. For accurate numerical results for the purpose of comparison with specific experimental results, one must resort to a full numerical solution of the Boltzmann integral equation which is well beyond the scope of the current work.

Our approximation does have the drawback that the Onsager relation $L_{TE} = -T L_{ET}$ is violated. This is a very generic feature of any two-relaxation-time version of the RTA. The physical reason is that the Boltzmann equation, in its total derivative form $df/dt=0$, is exactly the statement of conservation of particle number. As such, an uncontrolled approximate solution to the Boltzmann equation will generically lead to an uncontrolled non-conservation of particle number. Particle number conservation is assumed when proving this Onsager relation\cite{OnsagerRelations}, so there is no reason to expect that the Onsager relation will continue to hold for the approximate solution which violates this assumption. It so happens that a single-relaxation-time approximation does preserve the Onsager relation, but it will not typically lead to qualitatively correct results in the perturbative regime (see App. \ref{app:wf_single}).

Although our approximation violates the Onsager relations, this does not lead to qualitative changes in our conclusions as long as the qualitative behavior of $\sigma$ and $L_{TT}$ are correct. In particular, since the thermopower at low temperatures can only provide a negative correction to the approximation $\kappa \approx -L_{TT}$, the parametric suppression of the Lorenz number below $L_0$ that we have found can only be made more severe when the thermopower is accurately accounted for. Thus, the technical violation of the Onsager relation is an unimportant nuisance in our theory which we understand completely.  It arises simply from the fact that RTA by itself cannot provide an exact solution of the Boltzmann integral equation except in certain special situations.

In the main text, the numerical results are obtained from a schematic calculation for the thermal lifetime  Eq. \eqref{equ:tau_gamma}. This was done both for simplicity and for numerical stability. All of the results in the main text are in qualitative agreement with the RTA results in App.~\ref{app:phonon} in all asymptotic regimes, so the schematic numerical calculations are sufficient for our purposes.

\end{document}